\RequirePackage{fix-cm}
\documentclass[natbib]{svjour3}                     
\smartqed  
\usepackage{graphicx}
\usepackage{amsmath}
\usepackage[caption=false,font=footnotesize,labelfont=sf,textfont=sf]{subfig}
\usepackage{url}
\usepackage{booktabs}
\usepackage{multirow}
\usepackage[super]{nth}
\usepackage{soul} 
\usepackage{xcolor}

\newcommand{\ts}{\textsuperscript} 

\journalname{Social Network Analysis and Mining}
\begin{document}

\title{Exploring the effect of streamed social media data variations on social network analysis}
\subtitle{}

\titlerunning{Streamed social media data variations and SNA}        

\author{Derek Weber    \and
        Mehwish Nasim  \and
        Lewis Mitchell \and 
        Lucia Falzon
}

\authorrunning{D.C. Weber, M. Nasim, L. Mitchell \& L. Falzon} 

\institute{
    D.C. Weber 
    \at School of Computer Science, \\University of Adelaide / \\
        Defence Science and Technology Group\\
        Adelaide, South Australia, Australia\\
        ORCID ID: 0000-0003-3830-9014\\
        \email{derek.weber@\{adelaide.edu.au,dst.defence.gov.au\}}
    \and
    M. Nasim 
    \at Data61, CSIRO / \\
        Cyber Security Cooperative Research Centre / \\
        ARC Centre of Excellence for\\
        Mathematical and Statistical Frontiers\\
        Adelaide, South Australia, Australia\\
        ORCID ID: 0000-0003-0683-9125\\
        \email{mehwish.nasim@data61.csiro.au}
    \and
    L. Mitchell 
    \at School of Mathematical Sciences\\
        University of Adelaide\\
        Adelaide, South Australia, Australia\\
        ORCID ID: 0000-0001-8191-1997\\
        \email{lewis.mitchell@adelaide.edu.au}
    \and
    L. Falzon 
    \at School of Psychological Sciences\\
        University of Melbourne\\
        Melbourne, Victoria, Australia\\
        ORCID ID: 0000-0003-3134-4351\\
        \email{lucia.falzon@unimelb.edu.au}
}

\date{Received: date / Accepted: date}

\maketitle

\begin{abstract}
To study the effects of Online Social Network (OSN) activity on real-world offline events, researchers need access to OSN data,
the reliability of which has particular implications for social network analysis. 
This relates not only to the completeness of any collected dataset, but also to constructing meaningful social and information networks from them. 
In this multidisciplinary study, we consider the question of constructing traditional social networks from OSN data and then present several measurement case studies showing how variations in collected OSN data affects social network analyses.
To this end we developed a systematic comparison methodology, which we applied to five pairs of parallel datasets collected from Twitter in four case studies.
We found considerable differences in several of the datasets collected with different tools and that these variations significantly alter the results of subsequent analyses.

Our results lead to a set of guidelines for researchers planning to collect online data streams to infer social networks.

\end{abstract}

\section{Introduction}


Online activities can be associated with dramatic offline effects, such as voter fraud misinformation contributing to the 6 January 2020 riots and invasion of the US Capitol building in Washington DC \citep{Scott2021capitolriots}, COVID-19 misinformation leading to panic buying of toilet paper
~\citep{Yap2020loopaper}, online narratives incorrectly attributing Australia's ``Black Summer'' bushfires to arson amplifying attention to it in the media~\citep{WeberNFM2020arson}, and attempts to influence domestic and foreign politics~\citep{ratkiewicz2011,woolley2016autopower,morstatter2018alt,woolley2018us}. 
For researchers to successfully analyse online activity and provide advice about protection from such events, they must be able to reliably analyse data from online social networks (OSNs).

Social Network Analysis (SNA) facilitates exploration of social behaviours and processes. 
OSNs are often considered convenient proxies for offline social networks, because they seem to offer a wide range of data on a broad spectrum of individuals, their expressed opinions and inter-relationships. 
It is assumed that the social networks present on OSNs 
can inform the study of information dissemination and opinion formation, contributing to an understanding of offline community attitudes.
Though such claims are prevalent in the social media literature, there are serious questions about their validity due to an absence of SNA theory on online behaviour, the mapping between online and offline phenomena, and the repeatability of such studies. 
In particular, the issue of reliable data collection is fundamental. 
Collection of OSN data is often prone to inaccurate boundary specifications due to sampling issues, collection methodology choices, as well as platform  constraints. 

Previous work has considered the question of data reliability from a sampling perspective~\citep{morstatter2013sample,gonzalez2014assessing,JosephLC2014comparison,PaikLin2015}, biases~\citep{RuthsPfeffer2014,tromble2017we,pfeffer2018tampering,OlteanuCDK2019} and the danger of making invalid generalisations using ``big data'' approaches lacking nuanced interpretation of the data~\citep{lazer2014parable,tufekci2014big,falzon2017representataion,venturini2018actor}.
Analyses of incomplete networks exist~\citep{HolzmannAK2018}, but 
this paper specifically considers the questions of data reliability for SNA, 
considering not only the significance of 
online 
interactions to discover meaningful social networks, but also how sampling and boundary issues can complicate analyses of the networks constructed. Through an exploration of modelling and collection issues, and a 
measurement study examining the reliability of simultaneously collected, or \emph{parallel}, datasets, this multidisciplinary study addresses the following research questions:
\begin{itemize}
    \item \textit{To what extent do datasets obtained with social media collection tools differ, even when the tools are configured with the same search settings?}
    
    \item \textit{How do variations in collections affect the results of social network analyses?}
\end{itemize}
Our work makes the following contributions:

\begin{enumerate}
    \item Discussion of the challenges mapping OSN data to meaningful social and information networks;
    \item A methodology for systematic dataset comparison; 
    \item Recommendations for the use and evaluation of social media collection tools; and
    \item Five original social media datasets collected in parallel, and relevant analysis code\footnote{\url{https://github.com/weberdc/socmed_sna}}.
\end{enumerate}

This paper extends \cite{weber2020reliability} by including further detail in the methodology, incorporating further analyses and related visualisations and expanding the number of case studies considered from one to four. A discussion section also draws together observations from the case studies and explores ideas for a measure of dataset reliability.

\subsection{Paper Overview}

This paper continues with five sections: 1) challenges obtaining and modelling social networks from OSN data for SNA: 2) systematic parallel dataset comparison methodology; 3) results from using our methodology in a 
number of case studies; 4) discussion of our findings and exploration of the notion of a measure of reliability; and 5) recommendations for social media researchers and analysts, plus directions for future research.







\section{Social networks from social media} 

Using SNA to explore social behaviours and processes from OSN data presents many challenges. Most easily accessible OSN data consists of timestamped interactions, rather than details of long-standing relationships, which form the basis of SNA theory. Additionally, although interactions on different OSNs are superficially similar
, how they are implemented 
may subtly alter their interpretation. They offer a window onto online behaviour only, and any implications for offline relations and behaviour are unclear. Beyond 
modelling and reasoning with the data is the question of 
collection -- accessing the right data to construct meaningful social networks is challenging. 
OSNs 
provide a limited subset of their data through a variety of mechanisms, balancing privacy and 
competitive advantage with openness and transparency. 



\subsection{Interactions and relationships online} \label{sec:online_interactions}

SNA provides concepts and tools to model social relationships among actors. It is based on the premise that an actor's position in the network impacts their ability to access opportunities and resources and therefore allows us to understand social behaviours and processes in network terms \citep{borgatti2013analyzing}. Given the availability, nature and structure of much OSN data, the use of network-based techniques is a natural choice for the analysis of online social behaviour.

There is, however, an important distinction between the relatively stable, long-term relationships that are typically studied in SNA and the social connections among online actors~\citep{wasserman1994social,nasim2016inferring,borgatti2009network}. 
On social media, accounts can easily fulfil the role of actors, but precisely what constitutes a relationship is unclear. 
An obvious candidate is the \emph{friend} or \emph{follower} relationship common to most OSNs, 
but, due to 
how OSNs present their specific features to users,
each online community develops its own social relation culture. 
Therefore, such connections do not necessarily easily translate between OSNs. 
Is a Facebook \emph{friend}ship really the same as a \emph{follow} on Twitter, even if reciprocated? And how do each relate to offline friendships?

\begin{table}
    \centering
    \caption{Equivalent social media interaction primitives.}
    \label{tab:primitives}
    \resizebox{\columnwidth}{!}{%
        \begin{tabular}{@{}l|llllll@{}}
            \toprule
            OSN      & POST  & REPOST    & REPLY       & MENTION    & TAG        & LIKE \\
            \midrule
            Twitter  & tweet & retweet   & reply tweet & @mention   & \#hashtags & favourite \\
            Facebook & post  & share     & comment     & mention    & \#hashtag  & reactions \\
            Tumblr   & post  & repost    & comment     & @mention   & \#tag      & heart \\
            Reddit   & post  & crosspost & comment     & /u/mention & subreddit  & up/down vote \\
            \bottomrule
        \end{tabular}
    } 
\end{table}

OSNs offer ways to establish and maintain relations with others. 
This is done through interactions, many of which are common between OSNs, such as replying to the posts of others, mentioning others (causing the mentioning post to appear in the mentioned user's activity feed), using hashtags to reach broader communities, or sharing or reposting another's post to one's followers or friends. A sample of interactions with equivalents on different OSNs is offered in Table~\ref{tab:primitives}. (\textbf{N.B.} We distinguish interactions from follow or friending actions, which are used to define information flows and are persistent once created, requiring no further manipulation.) 
Specific interactions may be visible to different accounts, intentionally or incidentally (\emph{cf.} replying to one post versus using a hashtag). Exploration of these differences may lead to an understanding of the author's intent and  the identity of the intended audience. 
Is replying to a politician's Facebook post a way to connect directly with the politician, or is it a way to engage with the rest of the community replying to the post, either by specifically engaging with dialogue or merely signalling one's presence with a comment of support or dismay? 
A reply could be all of these things but, in particular, it is evidence of engagement at a particular time and indicates information flow between individuals~\citep{bagrow2019information}. Since most online interactions are directed towards a particular individual or group they offer an opportunity to study the flow of information and influence. 
On the other hand, although friend and follower connections may indicate community membership, they obscure the currency of that connection. 
Through their dynamic interactions, a user who \emph{liked} a Star Wars page ten years ago can be distinguished from one who not only \emph{liked} it, but posted original content to it on a monthly basis.
Therefore, we specifically focus on interactions rather than 
friend and follower relations in this study.


\subsection{Social Network Analysis theory}

Relationships between individuals in a social network may last for extended periods of time, vary in strength, and be based upon a variety of factors, not all of which are easily measurable. 
Because of the richness of the concept of social relationships, data collection is often a qualitative activity, involving directly surveying community members for their perceptions of their direct relations and then perhaps augmenting that data with observation data such as recorded interactions (e.g., meeting attendance, emails, phone calls). 
It is tempting to believe that this richness should be discoverable in the vast amount of interaction data available from OSNs, but there are issues to consider:
\begin{enumerate}
    \item links between social media accounts may vary in type and across OSNs --- it is unclear how they contribute to any particular relationship;
    \item what is observed online is only a partial record of interactions in a relationship, where interactions may occur via other OSNs or online media, or entirely offline; and
    \item collection strategies and OSN constraints may also hamper the ability to obtain a complete dataset.
\end{enumerate}
Although many interactions seem common across OSNs (e.g., a retweet on Twitter, a repost on Tumblr and a share on Facebook), nuances in how they are implemented and how data retrieved about them is modeled (beyond questions of semantics) may confound direct comparison. For example, a Twitter retweet refers directly to the original tweet, obscuring any chain of accounts through which it passed to the retweeter~\citep{RuthsPfeffer2014}. 
There are efforts to probabilistically regenerate such chains~\citep{DebateNightICWSM2018,gray2020bayesian}, but, in any case, is one account sharing the post further evidence of a relationship? What if it is reciprocated once, or three times? What if the reciprocation occurs only over some interval of time? These questions require careful consideration before SNA can be applied to OSN data. 

\subsection{Challenges obtaining OSN data}

Social media data is typically accessed via an OSN's Application Programming Interfaces (APIs), which place constraints on how true a picture researchers can form of any relationship. Via its API an OSN can control: \emph{how much data} is available, through rate limiting, biased or at least non-transparent sampling, and temporal constraints; \emph{what types of data} are available, through its data model; and \emph{how precisely data can be specified}, through its query syntax. 
Many OSNs offer commercial access, which provides more extensive access for a price, though use of such services in research raises questions of repeatability~\citep{RuthsPfeffer2014}. 
This is done to protect users' privacy but also to maintain competitive advantage. 
Researchers must often rely on the cost-free APIs, which present further issues. 
Twitter's 1\% Sample API 
has been found to provide highly similar samples to different clients and it is therefore unclear whether these are truly representative of Twitter traffic~\citep{JosephLC2014comparison,PaikLin2015}. 
If the samples were truly random, then they ought to be quite distinct, with only minimal overlap. 
Studying social media data therefore raises questions about the ``the coverage and representativeness''~\cite[p.17]{gonzalez2014assessing} of the sample obtained and how it therefore ``affects the networks of communication that can be reconstructed from the messages sampled''~\cite[p.17]{gonzalez2014assessing}. 

Empirical studies have compared the inconsistencies between collecting data from search and streaming APIs using the same or different lists of hashtags. 
Differences have been discovered between the free streaming API and the full (commercial) ``firehose'' API~\citep{morstatter2013sample}.
There is general agreement in the literature that the consistency of networks inferred from two streaming samples is greater when there is a high volume of tweets even when the list of hashtags is different~\citep{gonzalez2014assessing}.
More concerning is the ability to tamper with Twitter's sample API to insert messages~\citep{pfeffer2018tampering}, introducing unknown biases at this early stage of data collection~\citep{tromble2017we,OlteanuCDK2019}.

Taking a purely ``big data'' approach can also lead to inappropriate generalisations and conclusions~\citep{lazer2014parable,tufekci2014big}. This is well illustrated, for example, by the range of motivations behind retweeting behaviour including affirmation, sarcasm, disgust and disagreement~\citep{tufekci2014big}. Similarly, in the study of collective action, there are important social interactions that occur offline
~\citep{venturini2018actor}. 
Furthermore, relying solely on observable online behaviours risks overlooking passive consumers, resulting in underestimating the true extent to which social media 
can influence people~\citep{falzon2017representataion}.

OSN APIs provide data by streaming it live or through retrieval services, both of which make use of OSN-specific query syntaxes. Conceptually, therefore, there are two primary collection approaches to consider: 1) focusing on a user or users as seeds 
\citep[e.g.,][]{gruzd2011imagining,morstatter2018alt,KellerICWSM2017} using a snowball strategy~\citep{goodman1961} and 2) using keywords or filter terms  \citep[e.g.,][]{ratkiewicz2011,ferrara2017frelec,morstatter2018alt,woolley2018us,BessiFerrara2016,nasim2018real}. 
Focusing on seeds can reveal the flow of information within the communities around the seeds, while a keyword-based collection provides the ebb and flow of conversation related to a topic. 
These approaches can be combined, as exemplified by
~\citet{morstatter2018alt} in their study of the $2017$ German election: an initial keyword-based collection was conducted for eleven days to identify the most active accounts, the usernames of which were then used as keywords in a six week collection.

Once a reasonable dataset is obtained, there may be benefit in stripping junk content included by automated accounts such as bots~\citep{rise2016,botornot2016}. 
The question of whether to remove content from social bots (bots that actively pretend to be human) depends on the research question at hand; because humans are easily fooled by social bots~\citep{cresci2017,nasim2018real}, their contribution to discussions may still be valid (unlike, e.g., that of a sport score announcement bot). Several studies have examined how humans and bots interact, especially within political discussions~\citep{BessiFerrara2016,DebateNightICWSM2018,woolley2018us}.

So far, the following has been established: 
\begin{itemize}
    \item The OSN information used to form ties in social networks requires careful consideration to ensure meaningfulness; and
    \item Uncertainty regarding the completeness of OSN data 
    must be acccounted for.
\end{itemize}
We are now in a position to empirically examine the issue of repeatability, by comparing simultaneously retrieved collections.

\section{Methodology}

Our initial hypothesis was that if the same collection strategies were used at the same time, then each OSN would provide the same data, regardless of the collection tool used.
Consequently, social networks built from such data using the same methodology should be highly similar, in terms of both network and node level measurements. 
Our methodology consisted of these steps:
\begin{enumerate}
    \item Conduct simultaneous collections on an OSN using the same collection criteria with different tools.
    \item Compare statistics across datasets.
    \item Construct sample social networks from the data collected and compare network-level statistics.
    \item Compare the networks at the node level.
    \item Compare the networks at the cluster level.
\end{enumerate}

Examining the parallel datasets in each of these ways provides the opportunity for the analyst to develop a well-rounded understanding of the participants in an online discussion, their behaviour, how they relate to each other and the communities they form.

\subsection{Scope}

The scope of this work includes datasets obtained via streaming APIs filtered with keywords. Other collection styles may start with seed accounts, and collect their data and the data of accounts connected to them, either through interaction (e.g., via comments, replies or mentions) or via follower links. 
Such collections (especially follower networks) often require the collection of data that is prohibitive to obtain, is immediately out of date, and provides no real indication of strength of relationships, as discussed in Section~\ref{sec:online_interactions}. Additionally, in the absence of a domain-focused research question to inform the choice of seed accounts, no particular accounts would make sensible seeds, so here we rely on keyword-based collections.

\subsection{Data collection}


Twitter was chosen as the source OSN due to the availability of its data, the fact that the data it provides was thought to be highly regular~\citep{JosephLC2014comparison}, and because it has similar interaction primitives to other major OSNs. 
Twitter is also widely used in academia for research that makes predictions, in particular predictions about population-level events, behavioural patterns and information flows, such as studies of predicting social unrest~\citep{tuke2020pachinko} or misinformation~\citep{wu2016mining}. The validity of these predictions is fundamentally based on the consistency of the underlying (accessible) data. 
Two very different collection tools were chosen:

    \textbf{Twarc}\footnote{\url{https://github.com/DocNow/twarc}} is an open source library 
    which wraps 
    Twitter's API, and provided the baseline. 

    \textbf{RAPID} (Real-time Analytics Platform for Interactive Data Mining)~\citep{rapid2017} is a social media collection and data analysis platform for Twitter and Reddit. It enables filtering of OSN live streams, as well as dynamic \emph{topic tracking}, meaning it can update filter criteria in real time, adding terms popular in recent posts and removing unused ones. 


Both tools facilitate filtering Twitter's Standard live stream\footnote{\url{https://developer.twitter.com/en/docs/tweets/filter-realtime/overview}} with keywords,
providing datasets of tweets as JSON objects.

\subsection{Constructing social networks} \label{sec:net_cons}

A social network is constructed from dyads of pair-wise relations between nodes, which in our case are Twitter accounts. 
The node ties denote intermittent relations between accounts, inferred from observed interactions~\citep{nasim2016inferring,borgatti2009network}. 
Like any choice of knowledge representation, different networks can be constructed to address different research questions. For example, a network to study information flow could draw an arc from node A to B if account B retweets A's tweet (implying B has read and perhaps agreed with A's tweet); alternatively, the same interaction could be used to draw an arc from B to A if the relation is to imply an attribution of status or influence (A has influence because B has supported it through a retweet).  
Networks can be constructed based on direct or inferred relations, including retweeting, replying or mentioning, which we discuss below, or through the shared use of hashtags or URLs, reciprocation or minimum levels of interaction, or friend/follower connections. \citet{morstatter2018alt} constructed networks of accounts based on retweets and mentions to discover communities active during the 2017 German election, valuing mentions and retweets equally to mean one account reacting to another. URL sharing behaviour is often studied in the detection and classification of spam and political campaigns~\citep{CaoCLGC2015urlsh,Wu2018,Giglietto2020}. Some require more complex calculation such as linking accounts through their participation in detected events~\citep{nasim2018real}. Of course, applications for social network analysis exist outside the online sphere, e.g., in narrative analysis~\citep{edwards2018one}, and require similar considerations with regard to network design.
In the absence of clear alternative research questions, we will examine the social relationships implied by direct interactions and retweet networks (due to their frequency in the literature) and thus we will focus only on the three types of network construction discussed below.

Here, 
we consider three social networks 
built from 
interaction types 
common to many OSNs: `mention networks', `reply networks', and `retweet networks' (e.g., retweets are analogous to Facebook shares or Tumblr reposts, and replies analogous to comments on posts on Reddit, as shown in Table~\ref{tab:primitives}).
We define a social network, $G$=$(V,E)$, of accounts $u \in V$ linked by directed, weighted edges $(u_i,u_j) \in E$ based on the criteria below. 
\textbf{Mention Networks.} Twitter users can \emph{mention} one or more other users in a tweet. In a mention network, an edge $(u_i, u_j)$ exists iff $u_i$ mentions $u_j$ in a tweet, and the weight corresponds to the number of times $u_i$ has mentioned $u_j$.

\textbf{Reply networks.} A tweet can be a reply to one other tweet. In a reply network, an edge $(u_i, u_j)$ exists iff $u_i$ replies to a tweet by $u_j$, and the weight corresponds to the number of replies $u_i$ has made to $u_j$'s tweets. 


\textbf{Retweet networks.} A user can repost or `retweet' another's tweet on their own timeline (visible to their followers). 
Though retweets are not necessarily direct interactions~\citep{RuthsPfeffer2014}, they can be used to determine an account's reach,
and are widely used in the literature~\citep[e.g.,][]{vo2017,DebateNightICWSM2018,woolley2018us,morstatter2018alt}. 
In a retweet network, an edge $(u_i, u_j)$ exists iff $u_i$ retweets a tweet by $u_j$, and its weight corresponds to the number of $u_j$'s tweets $u_i$ has retweeted. 

\begin{figure}[!ht]
    \begin{minipage}[t]{\columnwidth}
        \subfloat[Mentions.\label{fig:qanda1_sample_mentions}]{%
            \includegraphics[height=0.155\textheight]{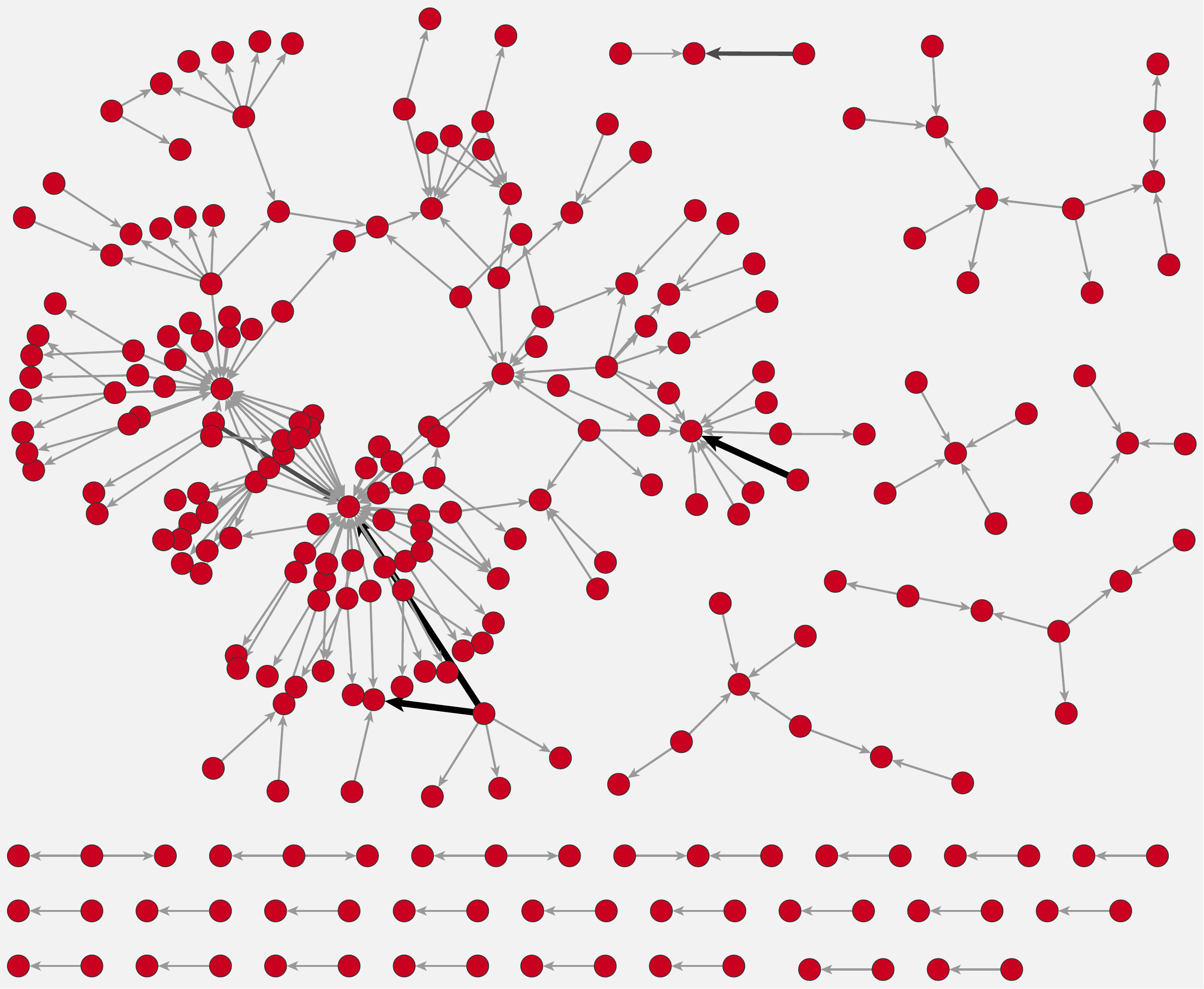}
        }
        \hfill
        \subfloat[Replies.\label{fig:qanda1_sample_replies}]{%
            \includegraphics[height=0.155\textheight]{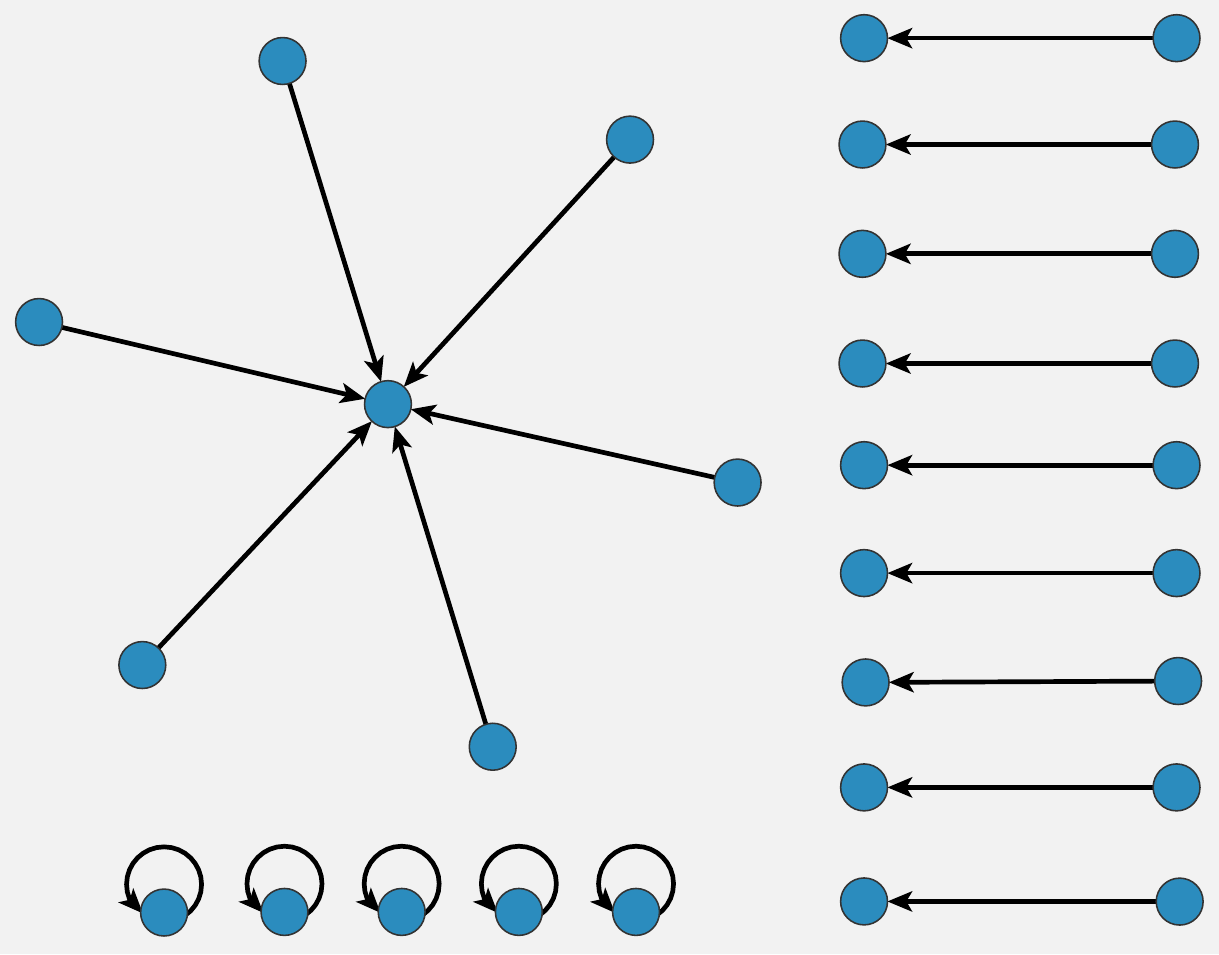}
        }
        \hfill
        \subfloat[Retweets.\label{fig:qanda1_sample_retweets}]{%
            \includegraphics[height=0.155\textheight]{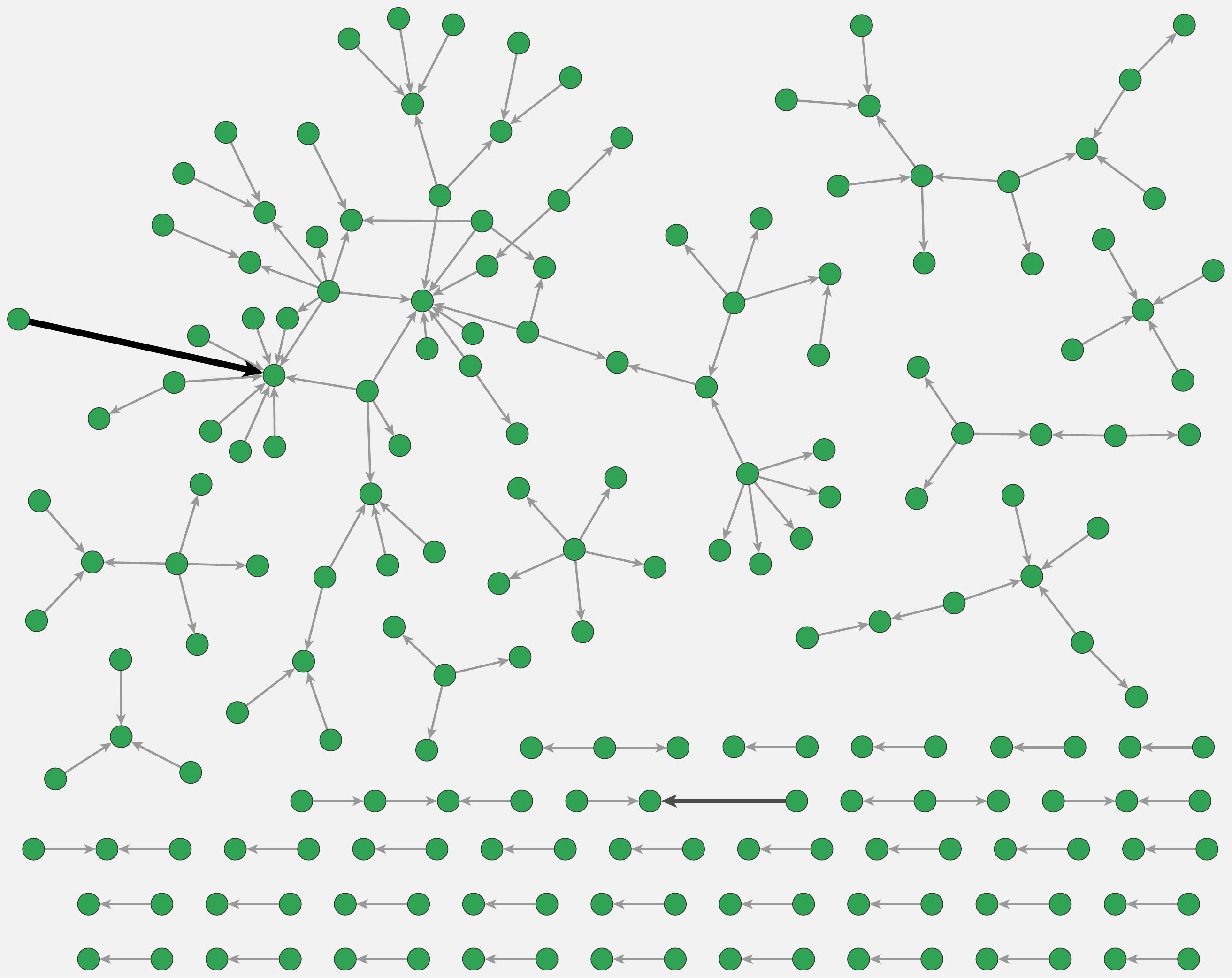}
        }
    \end{minipage}%
    \caption{Sample networks of accounts built from 5 minutes of Twitter data. 
    Nodes may appear in one or more networks, depending on their behaviour during the sampled period. The diagrams were constructed with \textit{visone} (https://visone.info).}
    \label{fig:qanda1_samples}
\end{figure}

Examining networks of the three types built from the same dataset, replies (Figure~\ref{fig:qanda1_sample_replies}) are the least common of the three interaction types, and all are dominated by a single large component. Mention networks exhibit relatively high cohesiveness. The similarity between retweets (Figure~\ref{fig:qanda1_sample_retweets}) and mentions (Figure~\ref{fig:qanda1_sample_mentions}) is because the data model of a retweet includes a mention of the retweeted account.

\color{black}






\subsection{Analyses} \label{sec:analyses}



At this point, comparative analysis can be applied to the parallel tweet datasets, initially by examining OSN-specific features and then the mention, reply and retweet networks constructed from them. When analysing these networks, it is relevant to note that SNA posits two important axioms on which most network measures are based: network structure affects collective outcomes; and positions within networks affect actor outcomes~\citep{robins2015doing}. Furthermore, we should expect minor differences in collections to be amplified in resulting social networks~\citep{HolzmannAK2018}.

\subsubsection{Dataset statistics}
To compare the parallel datasets we examined a number of features, their frequencies and several maximums. 
The first of these relate to the absolute count of the following features:
\begin{itemize}
    \item \textit{Tweets}: The number of tweets in the corpus.
    \item \textit{Accounts}: The number of unique accounts that posted tweets in the corpus (i.e., does not include those that were only mentioned or whose tweets were retweeted).
    \item \textit{Retweets}: The number of tweets which were native retweets (i.e., created by clicking the `retweet' button on the Twitter user interface, rather than manually typing in ``RT @original\_author: \emph{original text}'', which is another valid, though time consuming, way to post a retweet).
    \item \textit{Quotes}: The number of tweets which were quote tweets (non-native retweets, or retweets with comments).
    \item \textit{Replies}: The number of tweets which were replies, including replies to tweets outside of the corpus.
    \item \textit{URLs}: The number of tweets using URLs, the number of unique URLs used and the number of URL uses.
    \item \textit{Hashtags}: The number of tweets using hashtags, the number of unique hashtags used and the number of hashtag uses.
    \item \textit{Mentions}: The number of tweets containing mentions of other accounts, the number of unique mentioned accounts, and the number of mentions overall.
\end{itemize} 
The remainder relate to the highest values of the following features:
\begin{itemize}
    \item \textit{Tweeting account}: The most prolific account and the number of tweets they posted.
    \item \textit{Mentioned account}: The most mentioned account and the number of times they were mentioned.
    \item \textit{Retweeted tweet}: The most retweeted tweet and how often it was retweeted.
    \item \textit{Replied-to tweet}: The tweet with the most direct replies, and the number of those replies.
    \item \textit{Used hashtags}: The first and second most used hashtags, and the number of times they were used.
    \item \textit{URLs}: The most used URL, and the number of times it was used.
\end{itemize}

Based on these figures, we account for major discrepancies between the datasets, which can guide post-processing (e.g., spam filtering). Depending on the application domain, it may be appropriate to also consider comparing the distributions of particular features, rather than just their maximum values.

\subsubsection{Network statistics} The following network statistics are used to assess differences in the constructed networks: 
number of nodes, edges, average degree, density, mean edge weight, component count and the size and diameter of the largest, 
Louvain~\citep{blondel2008louvain} cluster count and the size of the largest, reciprocity, transitivity, and maximum k-cores. 
These measures provide us with an understanding of the `shape' of the networks in terms of how broad and dense they are and the strength of the connections within.

\subsubsection{Centrality values} 

Centrality measures offer a way to consider the importance of individual nodes within a network~\citep{newman2010networks}. The centrality measures considered here include: \emph{degree} centrality, indicating how many other nodes one node is directly linked to; \emph{betweenness} centrality, referring to the number of shortest paths in the network that a node is on and thus to what degree the node acts as a bridge between other nodes; \emph{closeness} centrality, which provides a sense of how topologically close a node is to the other nodes in a network; and \emph{eigenvector} centrality, which measures how connected a node is to other highly-connected nodes. Eigenvector centrality is often compared with Google's PageRank algorithm~\citep{brin1998anatomy}, which gives a measure of the importance of nodes (e.g., websites) based on references to them by other important nodes or by many nodes. The interested reader is referred to \citep{robins2015doing,wasserman1994social} for more detail.

Only centrality measures for mention and reply networks are considered, as edges in retweet networks are not direct interactions~\citep{RuthsPfeffer2014}.

Given the set of nodes in each corresponding pair of networks is not guaranteed to be identical, it is not possible to directly compare the centrality values of each node, so instead we rank the nodes in each network by the centrality values, take the top $1,000$ from each list, further constrain the lists to only the nodes common to both lists, and then compare the rankings. We initially compare the rankings visually using scatter plots
, where a node's rank in the first and second list is shown on the $x$ and $y$ axes, respectively.
A statistical measure of the similarity of the two rankings (of common nodes) is obtained with the Kendall $\tau$ coefficient, with Spearman's $\rho$ coefficient used as a confirmation measure. 
To classify the strength of the correlations, we followed the guidance of \citet[][p.175]{Dancey2011}, who posit that a coefficient of $0.0 - 0.1$ is uncorrelated, $0.11 - 0.4$ is weak, $0.41 - 0.7$ is moderate, $0.71 - 0.90$ is strong, and $0.91 - 1.0$ is perfect.

\subsubsection{Cluster comparison} The final step is to consider the clusters discoverable in the mention, reply and retweet networks and compare their membership. We first compare the distribution of the sizes of the twenty largest Louvain clusters~\citep{blondel2008louvain} visually. The Louvain method was chosen because it works well with large and small networks~\citep{yang2016comparative} and is used 
well known 
in the literature~\citep[e.g.,][]{morstatter2018alt,nasim2018real,Nizzoli2020}.

We use the Adjusted Rand index~\citep{HubertArabie1985adjrandindex} to compare membership. 
This considers two networks of the same nodes that have been partitioned into subsets. When considered in pairs, there are nodes that appear in the same subset in both partitions ($a$), and there are (many) pairs of nodes that do not appear in the same subsets in either partition ($b$), and the rest appear in the same subset in one of the partitions but not in the other. Defining the total of possible pairings of the $n$ nodes (\(\frac{n(n-1)}{2}\)) as $c$, the Rand index, $R$, is simply \(R = \frac{a + b}{c}\). The Adjusted Rand index (ARI) corrects for chance and provides a value in the range $[-1,1]$ where $0$ implies that the two partitions are random with respect to one another and~$1$ implies they are identical.

\section{Evaluation Case Studies}


Several case studies were conducted to evaluate the comparison methodology, the requirements for which developed progressively, each new case study's requirements informed by lessons from the previous.
The collections varied in the tools used for the parallel collections. As mentioned, Twarc was employed as a baseline, while RAPID was used with topic tracking enabled and disabled, and the tool Tweepy was used in only one case study as a second baseline.
The first case study consisted of two parallel Twitter datasets relating to an Australian panel discussion television programme with a prominent online community; the first datasets were collected over the running of the programme (4 hours) and the second covered the following day's discussion (15 hours), both employing RAPID's topic tracking feature to broaden the conversation. 
The second case study examined discussion surrounding Australian Rules Football over a longer period (3 days), without RAPID's topic tracking. 
The third also examined the online sports discussion, but over a longer time again (6 days) and only making use of RAPID with no topic tracking. 
The final case study incorporated a third tool to act as a further baseline and covered a regional but large election day, during which a significant amount of activity was expected. These conditions are summarised in Table~\ref{tab:collection_conditions}.

\begin{table}[t!h]
    \centering
    \caption{Summary of data collection conditions.}
    \label{tab:collection_conditions}
    \resizebox{\columnwidth}{!}{%
    \begin{tabular}{@{}clrlll@{}}
        \toprule
        Case Study & Collection  & Duration & Tool 1                    & Tool 2                      & Tool 3                 \\
        \cmidrule(r){1-1}\cmidrule(lr){2-2}\cmidrule(lr){3-3}\cmidrule{4-6}

        \multirow{2}{*}{1} & Q\&A Part 1 & 4 hours  & Twarc                     & RAPID (topic tracking)      & \multicolumn{1}{c}{---} \\
        & Q\&A Part 2 & 15 hours & Twarc                     & RAPID (topic tracking)      & \multicolumn{1}{c}{---} \\
                    \cmidrule(lr){2-2}\cmidrule(lr){3-3}\cmidrule{4-6}
        2 & AFL1        & 3 days   & Twarc                     & RAPID (no topic tracking)   & \multicolumn{1}{c}{---} \\
                    \cmidrule(lr){2-2}\cmidrule(lr){3-3}\cmidrule{4-6}
        3 & AFL2        & 6 days   & RAPID (no topic tracking) & RAPID (no topic tracking)   & \multicolumn{1}{c}{---} \\
                    \cmidrule(lr){2-2}\cmidrule(lr){3-3}\cmidrule{4-6}
        4 & Election    & 1 day    & Twarc                     & RAPID (with topic tracking) & Tweepy                  \\
        \bottomrule
    \end{tabular}
    } 
\end{table}

A summary of the corpora collected is presented in Table~\ref{tab:all_collection_stats}. As noted above, when topic tracking was employed with RAPID, some of the tweets it collected did not contain any of the initial keywords. These datasets are given the label `RAPID-E'. Prior to comparison with the corresponding Twarc datasets, the RAPID-E datasets were filtered to retain only tweets containing at least one of the original keywords. 
The AFL2 case study used RAPID with no expansion with two sets of Twitter credentials simultaneously; in this case the datasets are labelled `RAPID1' and `RAPID2'.
A third collection tool, Tweepy\footnote{Tweepy is another open source library which provides a thin wrapper around the TwitterAPI: \url{https://github.com/tweepy/tweepy}.}, was included in the Election Day case study to act as a second baseline.

\begin{table}[t!h]
    \centering
    \caption{Summary statistics for the datasets used in this paper.}
    \label{tab:all_collection_stats}
    \begin{tabular}{@{}llrr@{}}
        \toprule
                      & Dataset & Tweets & Accounts \\
        \midrule
        Q\&A Part 1    & Twarc   & 27,389 &    7,057 \\
                      & RAPID   & 15,930 &    4,970 \\
                      & RAPID-E & 17,675 &    5,547 \\
        \midrule
        Q\&A Part 2    & Twarc   & 15,490 & 5,799 \\
                      & RAPID   & 11,719 & 4,708 \\
                      & RAPID-E & 23,583 &    8,854 \\
        \midrule
        AFL1          & Twarc   & 44,470  &   16,821  \\
                      & RAPID   & 21,799  &   11,573  \\
        \midrule
        AFL2          & RAPID1  & 30,103 &   14,231 \\
                      & RAPID2  & 30,115 & 14,232  \\
        \midrule
        Election Day  & Twarc   & 39,297 &   10,860 \\
                      & Tweepy  & 36,172 & 10,242  \\
                      & RAPID   & 39,556 & 10,893 \\
                      & RAPID-E & 46,526 & 12,696 \\
        \bottomrule
    \end{tabular}
\end{table}


\subsection{Case Study 1: Q\&A, \texttt{\#qanda} and the effect of topic tracking}


Initially, to obtain a moderately active portion of activity, we collected data from Twitter's Standard live stream\footnote{All data were collected, stored, processed and analysed according to two ethics protocols 
\#170316 and H-2018-045, approved by the University of Adelaide's 
human research and ethics committee.} relevant to an Australian television panel show, Q\&A, that invites its viewers to participate in the discussion live\footnote{The Australian Broadcasting Commission's ``Q\&A'' observes \#QandA.}. 
A particular broadcast in $2018$ was chosen due to the expectation of high levels of activity given the planned discussion topic. 
As a result, the filter keywords used were 
`qanda'\footnote{The `\#' was omitted to catch mentions of `@qanda', the programme's Twitter account.} and two terms 
that identified 
a panel member (available on request). 
We collected two parallel datasets: 

    \textbf{Q\&A Part 1:} Four hours starting 
    $30$ minutes before the hour-long programme, 
    to allow for contributions from the country's major timezones; and
    
    \textbf{Q\&A Part 2:} From 6am to 9pm the following day, capturing 
    further related online discussions. 

Twarc acted as the baseline collection as it provides direct access to Twitter's API, 
while 
RAPID was configured to use topic tracking via \emph{co-occurrence keyword expansion}~\citep{rapid2017}, meaning it would progressively add keywords to the original set if they appeared sufficiently frequently (five times in ten minutes).
Expanded datasets such as these are referred to as `RAPID-E'; it was filtered back to just the tweets containing the original keywords and labelled `RAPID' to enable fair comparison with the `Twarc' dataset. 
We expected the moderate activity observed would not breach rate limits, and thus RAPID should capture all tweets captured by Twarc. This was not the case. 
\begin{table}[t]
    \centering
    \resizebox{\columnwidth}{!}{%
        \begin{tabular}{@{}ll|r|rr|rr|r|rr@{}}
        \toprule
                      & Dataset & \multicolumn{1}{c}{All} & \multicolumn{2}{c}{Unique} & \multicolumn{2}{c}{Retweets} & \multicolumn{1}{c}{All} & \multicolumn{2}{c}{Unique} \\ 
                      &         & \multicolumn{1}{c}{Tweets} & \multicolumn{2}{c}{Tweets} & \multicolumn{2}{c}{} & \multicolumn{1}{c}{Accounts} & \multicolumn{2}{c}{Accounts} \\
        \midrule
        Q\&A Part 1   & Twarc   & 27,389 & 11,481 & (41.9\%) & 14,191   & (51.8\%) & 7,057    & 2,090    & (29.6\%) \\
                      \cmidrule{2-10}
        (20:00-00:00) & RAPID   & 15,930 & 22     & (0.1\%)  & 8,744    & (54.9\%) & 4,970    & 3        & (0.1\%)  \\
                      & RAPID-E & 17,675 & 1,767  & (10.0\%) & 9,767    & (55.3\%) & 5,547    & 527      & (9.5\%)  \\
        \midrule
        Q\&A Part 2   & Twarc   & 15,490 & 4,089  & (26.4\%) & 10,988   & (70.9\%) & 5,799    & 1,128    & (19.5\%) \\
                      \cmidrule{2-10}
        (06:00-21:00) & RAPID   & 11,719 & 318    & (2.7\%)  & 8,051    & (68.7\%) & 4,708    & 37       & (0.8\%)  \\
                      & RAPID-E & 23,583 & 12,180 & (51.6\%) & 13,679   & (58.0\%) & 8,854    & 4,007    & (45.3\%) \\
        \bottomrule
        \end{tabular}%
    }
    \caption{Summary statistics for the Q\&A Parts 1 and 2 datasets.}
    \label{tab:qanda_collection_stats}
\end{table}



\subsubsection{Comparison of collection statistics}

The first striking difference between the datasets was the number of tweets collected and the effect on the number of contributors (Table~\ref{tab:qanda_collection_stats}). 
RAPID collected fewer tweets by fewer accounts, but the datasets were close to subsets of the Twarc datasets. 
Between 26\% and 42\% of the tweets collected by Twarc were missed by RAPID, but the proportion of retweets in each part is similar (52\% and 55\% for Part 1 and 69\% and 71\% for Part 2). 
In both parts, very few accounts appear in only the RAPID collections. 
Discussions with RAPID's developers revealed it dumps tweets that miss the filter terms from the textual parts of tweets (e.g., the body, the author's screen name, and the author's profile description). The extra tweets RAPID collected were relevant and in English\footnote{Sometimes short or obscure filter terms, like `qanda', have meanings in non-target languages.} (based on manual inspection) but posted by different accounts (unique to RAPID-E). Of the tweets that RAPID collected which contained the keywords, they were posted by almost the same accounts as Twarc, but simply did not contain the same tweets.

The benefit of topic tracking via keyword expansion is yet to be strongly evaluated, but this study indicates there are benefits (relevant tweets that omit the original filter terms are picked up once related terms are added) as well as costs 
(tweets that include the original filter terms but are not collected). 
RAPID's expansion strategies are modifiable to optimise data collection, however we chose not to make use of this capability to prevent obscuring the current comparative study. 
The rest of this analysis explores how much of a difference the keyword expansion makes with regard to SNA.

\begin{table}[ht]
    \centering
    \caption{Detailed statistics of Q\&A Parts 1 and 2.}
    \label{tab:more_qanda_collection_stats}
    \resizebox{\columnwidth}{!}{%
        \begin{tabular}{@{}l|rr|rr|@{}}
            \toprule
                                               & \multicolumn{2}{c}{Q\&A Part 1} & \multicolumn{2}{c}{Q\&A Part 2} \\ 
                                               &         RAPID &         Twarc &         RAPID &         Twarc \\
            \midrule
            Tweets                             &        15,930 &        27,389 &        11,719 &        15,490 \\
            \midrule
            Quotes                             &           325 &         1,203 &           498 &         1,232 \\
            Replies                            &         1,446 &         2,067 &         1,715 &         1,731 \\
            Tweets with hashtags               &        10,043 &        15,591 &         3,912 &         3,961 \\
            Tweets with URLs                   &         2,470 &         4,029 &         3,106 &         4,074 \\
            Most prolific account              & Account $a_1$ & Account $a_1$ & Account $a_2$ & Account $a_3$ \\
            Tweets by most prolific account    &           103 &           146 &            57 &            68 \\
            Most retweeted tweet               &   Tweet $t_1$ &   Tweet $t_1$ &   Tweet $t_2$ &   Tweet $t_2$ \\
            Most retweeted tweet count         &           260 &           288 &           385 &           385 \\
            Most replied to tweet              &   Tweet $t_3$ &   Tweet $t_3$ &   Tweet $t_4$ &   Tweet $t_4$ \\
            Most replied to tweet count        &            55 &           121 &            58 &            58 \\
            Tweets with mentions               &        11,314 &        18,253 &        10,472 &        13,514 \\
            Most mentioned account             & Account $a_4$ & Account $a_4$ & Account $a_4$ & Account $a_4$ \\
            Mentions of most mentioned account &         2,883 &         3,853 &         2,753 &         2,752 \\
            Hashtags uses                      &        15,700 &        23,557 &         7,672 &         7,862 \\
            Unique hashtags                    &         1,015 &         1,438 &           960 &         1,082 \\
            Most used hashtag                  &       \#qanda &       \#qanda &       \#qanda &       \#qanda \\
            Uses of most used hashtag          &        10,065 &        15,644 &         2,545 &         2,549 \\
            Next most used hashtag             &      \#auspol &      \#auspol &      \#auspol &      \#auspol \\
            Uses of next most used hashtag     &         1,381 &         2,103 &         1,652 &         1,349 \\
            URLs uses                          &           913 &         1,650 &         1,602 &         2,411 \\
            Unique URLs                        &           399 &           560 &           658 &           790 \\
            Most used URL & \url{http://wp.me/p2WW3S-Gg} & \url{http://wp.me/p2WW3S-Gg} & Tweet $t_5$ URL & Tweet $t_6$ URL \\ 
            Uses of most used URL              &            49 &           128 &            71 &            81 \\
            \bottomrule
        \end{tabular}
    } 
\end{table}

\begin{table}[ht!]
    \centering
    \caption{The top ten most used hashtags in the Q\&A datasets (ignoring case and anonymising names).}
    \label{tab:qanda_top_hashtags}
    \resizebox{\columnwidth}{!}{%
        \begin{tabular}{@{}lrlr|lrlr@{}}
        \toprule
        \multicolumn{4}{c}{Q\&A Part 1}                        & \multicolumn{4}{c}{Q\&A Part 2}                         \\ 
        \multicolumn{2}{c}{RAPID}  & \multicolumn{2}{c}{Twarc} & \multicolumn{2}{c}{RAPID}  & \multicolumn{2}{c}{Twarc}  \\
        \multicolumn{2}{c}{\emph{15,930 tweets}} & \multicolumn{2}{c}{\emph{27,389 tweets}} & \multicolumn{2}{c}{\emph{11,719 tweets}} & \multicolumn{2}{c}{\emph{15,490 tweets}} \\
        Hashtag                    & Count  & Hashtag          & Count  & Hashtag           & Count & Hashtag           & Count \\
        \midrule
        qanda                      & 10,065 & qanda            & 15,644 & qanda             & 2,545 & qanda             & 2,549 \\
        auspol                     & 1,381  & auspol           & 2,103  & auspol            & 1,652 & auspol            & 1,349 \\
        ulurustatement             & 179    & nbn              & 223    & Surname of $a_4$  & 179   & Surname of $a_4$  & 179   \\
        nbn                        & 178    & ulurustatement   & 187    & nbn               & 135   & nbn               & 133   \\
        Surname of $a_4$           & 137    & Surname of $a_4$ & 179    & breaking          & 85    & ulurustatement    & 73    \\
        marriageequality           & 125    & marriageequality & 145    & ulurustatement    & 72    & pmlive            & 71    \\
        felizjueves                & 114    & felizjueves      & 128    & qldpol            & 65    & qldpol            & 64    \\
        climate                    & 73     & ssm              & 80     & nswpol            & 65    & nswpol            & 63    \\
        8kasımdünyadelilergünü     & 61     & climate          & 77     & pmlive            & 64    & marriageequality  & 53    \\
        ssm                        & 60     & libspill         & 76     & marriageequality  & 53    & springst          & 49    \\        \bottomrule
        \end{tabular}%
    }
\end{table}

Table~\ref{tab:more_qanda_collection_stats} reveals that although feature counts vary significantly, many of the most common values are the same (e.g., most retweeted tweet, most mentioned account, most used hashtags). 
Many are approximately proportional to corpus size (Twarc is 1.72 and 1.32 times larger than RAPID for Parts 1 and 2, respectively), but with notable exceptions and no apparent pattern. 
Some values are remarkably similar, despite the size of the corpora they arise from being so different. 
For example, Twarc picked up nearly $8,000$ more hashtag uses than RAPID in Part 1, but fewer than $200$ more in Part 2. 
Notably, although the most prolific account is different in Part~2, the most mentioned account is the same for both Parts~1 and~2, potentially implying that account has had similarly high influence in both parallel datasets. Furthermore, both datasets shared almost all the same top ten hashtags, though in different orders (see Table~\ref{tab:qanda_top_hashtags}).  
Approximately $5,000$ of the extra hashtag uses are of `\#qanda'. In Part 2, again, the top ten hashtags are nearly the same, but this time the usage counts are similar, except for `\#auspol' being used $22\%$ more often in RAPID ($1,652$ times compared with $1,349$), which would account for the overall difference of $190$ uses when combined with the noise of lesser used hashtags. The most used URL in Part~1 is a shortened form of a link to a political party policy comparison resource prepared by an account prominent in the \#auspol Twitter discussion\footnote{\url{https://otiose94.wordpress.com/2015/05/30/nett_news-by-otiose94/}}. 
In the longer collection, the most prominent URL is overtaken by retweets, one by @QandA (Tweet $t_5$) and one by @SkyNewsAust, an official news media account (Tweet $t_6$).

\begin{figure}[!ht]
    \begin{minipage}[t]{\columnwidth}
        \subfloat[Q\&A Part 1.\label{fig:qanda1_timeline}]{
            \includegraphics[width=0.49\columnwidth]{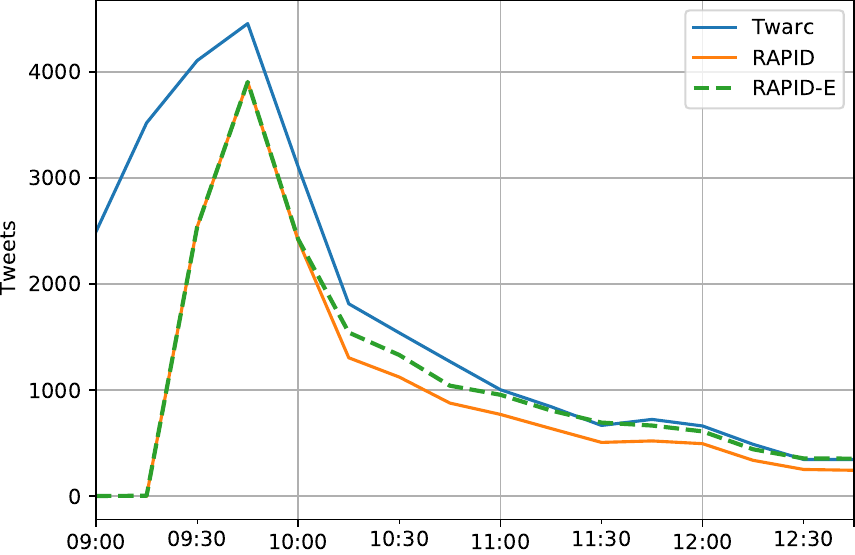}
        }
        \hfill
        \subfloat[Q\&A Part 2.\label{fig:qanda2_timeline}]{
            \includegraphics[width=0.49\columnwidth]{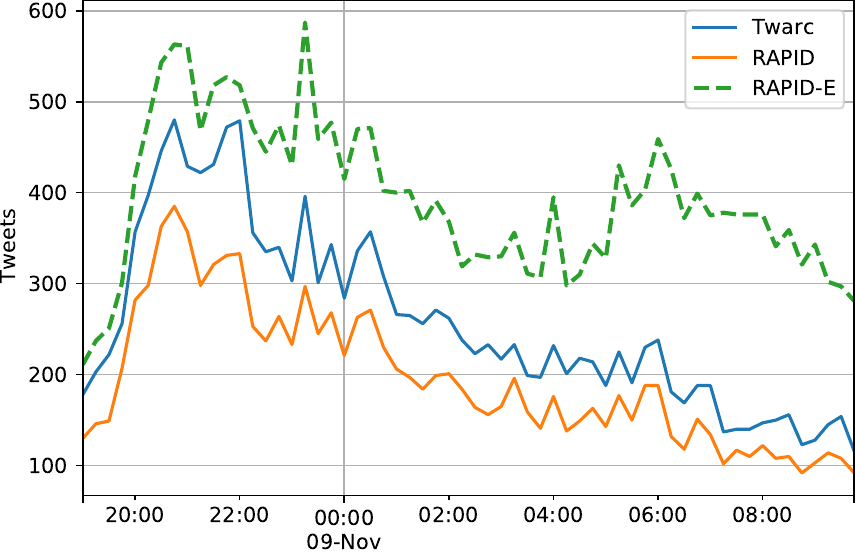}
        }
    \end{minipage}
    \caption{Twitter activity in the Q\&A Parts 1 and 2 dataset over time (in 15 minute blocks).}
    \label{fig:qanda_timelines}
\end{figure}

\begin{figure}[!ht]
    \begin{minipage}[t]{\columnwidth}
        \subfloat[From the first 5000 tweets.\label{fig:qanda2_first_5000_terms_wc}]{
            \includegraphics[width=0.32\columnwidth]{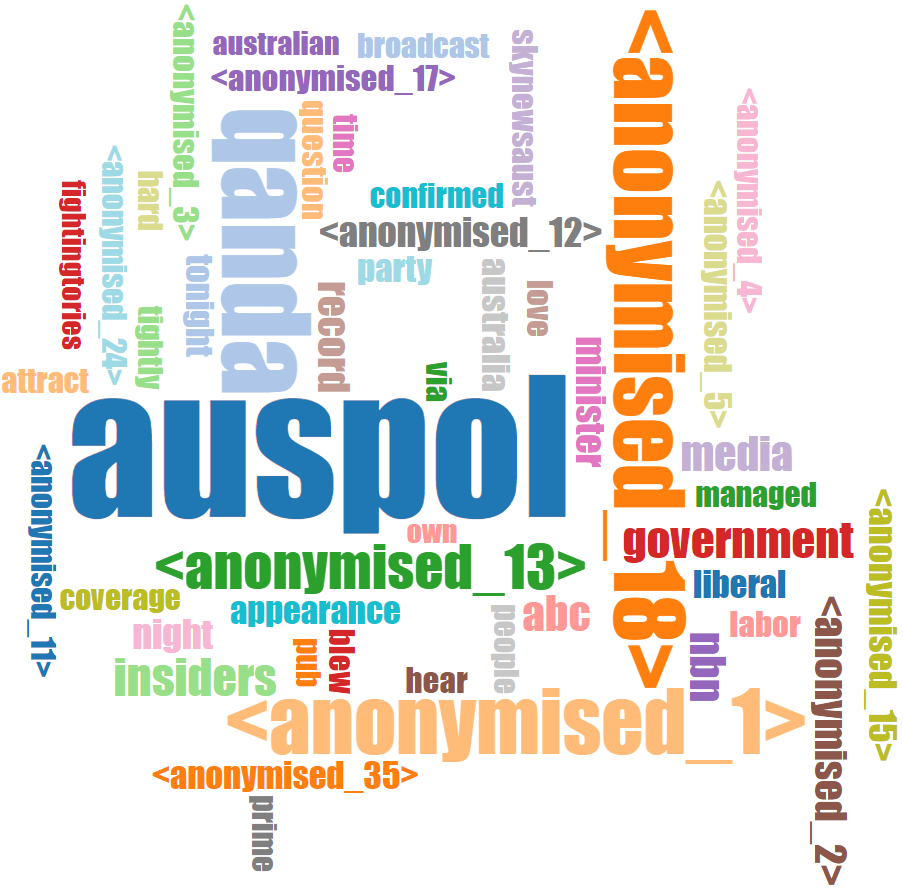}
        }
        \hfill
        \subfloat[From the last 5000 tweets.\label{fig:qanda2_last_5000_terms_wc}]{
            \includegraphics[width=0.32\columnwidth]{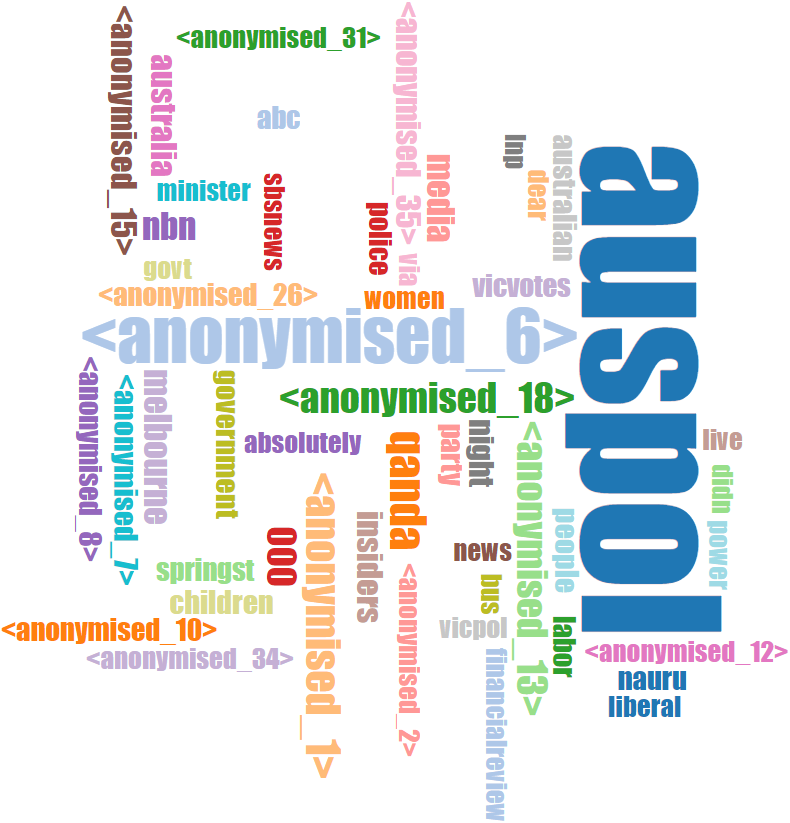}
        }
        \subfloat[Unique to the last 5000 tweets.\label{fig:qanda2_last_5000_unique_terms_wc}]{
            \includegraphics[width=0.32\columnwidth]{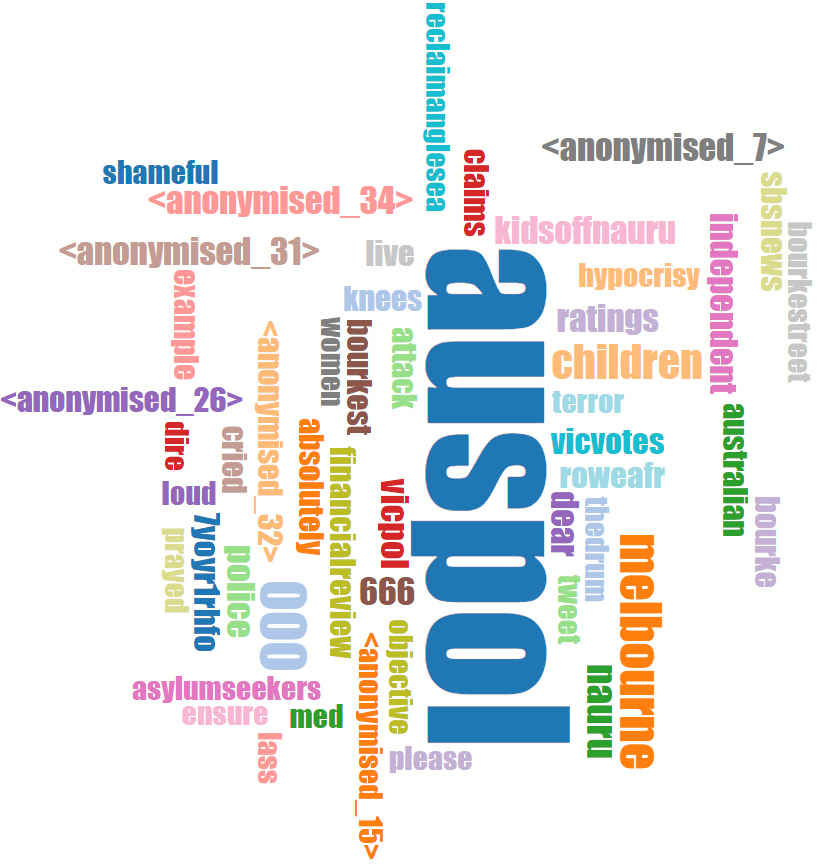}
        }
    \end{minipage}
    \caption{Word clouds of the $50$ most used terms (anonymised) in the first and last $5,000$ tweets of the Q\&A Part 2 RAPID-E dataset, and the top $50$ terms unique to the last $5,000$ tweets.}
    \label{fig:qanda2_early_late_term_comparisons}
\end{figure}

Moving beyond the bare statistics, the timelines shown in Figures~\ref{fig:qanda1_timeline} and~\ref{fig:qanda2_timeline} show the clear differences in tweets retrieved. Though the Twarc and pared back RAPID timelines appear at least proportionately similar, it is firstly notable that the RAPID-E dataset captured so much less data in Part 1 (Figure~\ref{fig:qanda1_timeline}) and so much more in Part 2 (Figure~\ref{fig:qanda2_timeline}), particularly from approximately 4~a.m. onwards (UTC). One possible explanation for this is that the discussion on the night of the episode was far more directly focused on the episode themes and had less opportunity to drift to other issues, especially while informed and guided by what was being broadcast at the time. In contrast, those discussing the episode the following day would have had more opportunity to broaden the discussion to other topics, and RAPID's topic tracking attended to that, apparently at the cost of tweets matching the exact filter terms. Word clouds of the terms drawn from the first and last $5,000$ tweets of the RAPID-E dataset appear to offer mild support for this (Figure~\ref{fig:qanda2_early_late_term_comparisons}). Terms are sized according to their frequency. The discussion across the day focuses on the \texttt{\#auspol} hashtag, but \texttt{\#qanda} is more prominent early on. Mentions of anonymised IDs 1 and 18 are prominent early but shift to ID 6 later. All of these IDS refer to the same individual\footnote{Variants of this individual's name were used as filter terms.}, but by Twitter handle and first name early on and by surname later in the day. Figure~\ref{fig:qanda2_last_5000_unique_terms_wc}, showing the top terms unique to the evening discussion, indicates that the discussion shifts to humanitarian concerns, perhaps due to events of the day. The early discussion (Figure~\ref{fig:qanda2_first_5000_terms_wc}) seems to mention individuals much more than later, as indicated by the greater size of anonymised IDs. This fact alone implies that the early discussion was focused more directly on the the Q\&A episode, as the topics it covered related to particular relationships and events involving those individuals.

The second notable feature is that the RAPID tool appeared to miss many of the available tweets in the first half an hour of the Part~1 collection. RAPID-E's first half hour includes only six tweets, the first of which was at 9~a.m. (UTC), while RAPID's only includes four tweets, the first of which was at 9:15~a.m. It is unclear why the tool missed the tweets that Twarc captured, but a discrepancy such as this suggests it was not by design. The reason that the RAPID-E included tweets without the key terms early on in the specified timeframe is that the collection was running prior to the cut-off at 9~a.m. (UTC), tracking topics while it ran, as a `burn-in' period, and we have extracted just these specific periods (UTC 0900 to 1300, and UTC 1900 to 1000 the next day) to study, post collection.

\subsubsection{Comparison of network statistics}

\begin{figure}[!ht]
    \centering
    \includegraphics[width=0.99\columnwidth]{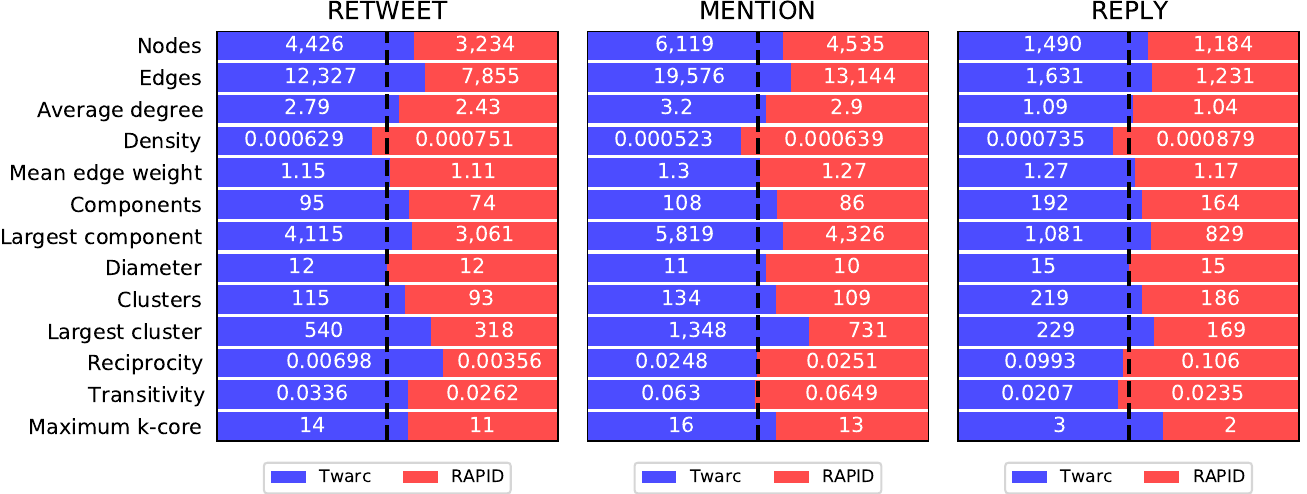}
    \caption{The proportional balance between Twarc and RAPID statistics of the retweet, mention and reply networks built from the Q\&A Part 1 datasets.}
    \label{fig:qanda1_graph_stats_barh}
\end{figure}
\begin{figure}[!ht]
    \centering
    \includegraphics[width=0.99\columnwidth]{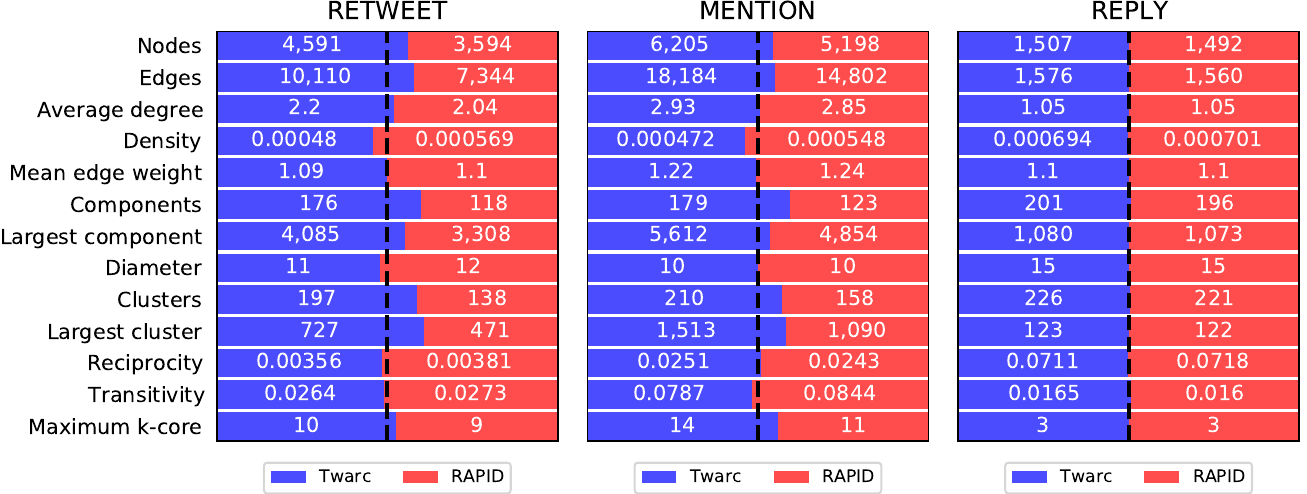}
    \caption{The proportional balance between Twarc and RAPID statistics of the retweet, mention and reply networks built from the Q\&A Part 2 datasets.}
    \label{fig:qanda2_graph_stats_barh}
\end{figure}

Given the differences in datasets, we expect differences in the derived social networks (Tables \ref{tab:qanda1_graph_stats} and \ref{tab:qanda2_graph_stats})~\citep{HolzmannAK2018}. 
We also present 
the proportional balance between each dataset's statistics in Figures~\ref{fig:qanda1_graph_stats_barh} and~\ref{fig:qanda2_graph_stats_barh}. 
Each network is dominated by a single large component, comprising over 90\% of nodes in the retweet and mention networks, and around 70\% in the reply networks.
The distributions of component sizes appear to follow a power law, 
resulting in 
corresponding high numbers of detected clusters.


\begin{table}[t]
    \centering
    \caption{Q\&A Part 1 network statistics.}
    \label{tab:qanda1_graph_stats}
    \resizebox{\columnwidth}{!}{%
    \begin{tabular}{@{}l|rr|rr|rr@{}}
        \toprule
                          & \multicolumn{2}{c}{RETWEET} & \multicolumn{2}{c}{MENTION} & \multicolumn{2}{c}{REPLY} \\ 
                          & RAPID       & Twarc         & RAPID       & Twarc         & RAPID       & Twarc       \\
        \midrule 
        Nodes             & 3,234       & 4,426         & 4,535       & 6,119         & 1,184       & 1,490       \\
        Edges             & 7,855       & 12,327        & 13,144      & 19,576        & 1,231       & 1,631       \\
        Average degree    & 2.429       & 2.785         & 2.898       & 3.199         & 1.040       & 1.095       \\
        Density           & 0.001       & 0.001         & 0.001       & 0.001         & 0.001       & 0.001       \\
        Mean edge weight  & 1.113       & 1.151         & 1.268       & 1.300         & 1.175       & 1.267       \\
        Components        & 74          & 95            & 86          & 108           & 164         & 192         \\
        Largest component & 3,061       & 4,115         & 4,326       & 5,819         & 829         & 1,081       \\
        - Diameter        & 12          & 12            & 10          & 11            & 15          & 15          \\
        Clusters          & 93          & 115           & 109         & 134           & 186         & 219         \\
        Largest cluster   & 318         & 540           & 731         & 1,348         & 169         & 229         \\
        Reciprocity       & 0.004       & 0.007         & 0.025       & 0.025         & 0.106       & 0.099       \\
        Transitivity      & 0.026       & 0.034         & 0.065       & 0.063         & 0.024       & 0.021       \\
        Maximum k-core    & 11          & 14            & 13          & 16            & 2           & 3           \\ 
        \bottomrule
    \end{tabular}
    } 
\end{table}

\begin{table}[!ht]
    \centering
    \caption{Q\&A Part 2 network statistics.}
    \label{tab:qanda2_graph_stats}
    \resizebox{\columnwidth}{!}{%
    \begin{tabular}{@{}l|rr|rr|rr@{}}
        \toprule
                          & \multicolumn{2}{c}{RETWEET} & \multicolumn{2}{c}{MENTION} & \multicolumn{2}{c}{REPLY} \\ 
                          & RAPID       & Twarc         & RAPID       & Twarc         & RAPID       & Twarc       \\
        \midrule 
        Nodes             & 3,594       & 4,591         & 5,198       & 6,205         & 1,492       & 1,507       \\
        Edges             & 7,344       & 10,110        & 14,802      & 18,184        & 1,560       & 1,576       \\
        Average degree    & 2.043       & 2.202         & 2.848       & 2.931         & 1.046       & 1.046       \\
        Density           & 0.001       & 0.000         & 0.001       & 0.000         & 0.001       & 0.001       \\
        Mean edge weight  & 1.096       & 1.087         & 1.245       & 1.222         & 1.099       & 1.098       \\
        Components        & 118         & 176           & 123         & 179           & 196         & 201         \\
        Largest component & 3,308       & 4,085         & 4,854       & 5,612         & 1,073       & 1,080       \\
        - Diameter        & 12          & 11            & 10          & 10            & 15          & 15          \\
        Clusters          & 138         & 197           & 158         & 210           & 221         & 226         \\
        Largest cluster   & 471         & 727           & 1,090       & 1,513         & 122         & 123         \\
        Reciprocity       & 0.004       & 0.004         & 0.024       & 0.025         & 0.072       & 0.071       \\
        Transitivity      & 0.027       & 0.026         & 0.084       & 0.079         & 0.016       & 0.016       \\
        Maxmium k-core    & 9           & 10            & 11          & 14            & 3           & 3           \\ 
        \bottomrule
    \end{tabular}
    } 
\end{table}


Network structure statistics like density, diameter (of the largest component in disconnected networks), reciprocity and transitivity may offer insight into social behaviours such as influence and information gathering. 
The high component counts in all networks lead to low densities and correspondingly low transitivities, as the potential number of triads is limited by the connectivity of nodes. 
That said, the largest components were consistently larger in the Twarc datasets, but the diameters of the corresponding largest components from each dataset were remarkably similar, implying that the extra nodes and edges were in the components' centres rather than on the periphery. 
This increase in internal structures 
improves connectivity and therefore the number of nodes to which any one node could pass information (and therefore influence) or, at least, reduces the length of paths between nodes so information can pass more quickly. 
The similarities in transitivity imply the increase may not be significant, however, with networks of these sizes. Reciprocity values may provide insight into information gathering, which often relies on patterns of to-and-fro communication as a person asks a question and others respond. Interestingly, the only significant difference in reciprocity is in the Part 1 retweet networks, with the Twarc dataset having a reciprocity nearly double that of the RAPID dataset (though still small). The Twarc dataset includes 60\% more retweets than the corresponding RAPID dataset and 40\% more accounts (Table~\ref{tab:qanda_collection_stats}), 
which may account for the discrepancy. 
Given the network sizes, the reciprocity values indicate low degrees of 
conversation, mostly in the reply networks. 
Interestingly, mean edge weights are very low ($1.3$ at most), implying that most interactions between accounts in all networks happen only once, despite these being corpora of issue-based discussions.

The proportional statistical differences between the corresponding datasets are highlighted in Figure~\ref{fig:qanda1_graph_stats_barh} for Part~1 and Figure~\ref{fig:qanda2_graph_stats_barh} for Part~2. Part~1's Twarc networks were larger, both in nodes and edges, but less dense, than the RAPID ones, and the largest component in each network is larger by a significant proportion of the extra nodes (it is not clear what portion of the extra nodes are members of the largest components, however). An increase in components also led to an corresponding increase in detected clusters, and an increase in the size of the largest detected cluster. As mentioned earlier, the increase in internal structures leads to a higher maximum k-core value. Though the proportional differences in reciprocity in the retweet networks are high, the values themselves remain low. Part~2's reply networks are remarkably similar despite the Twarc dataset having $26\%$ more tweets. The differences in Part~2's retweet and mention networks are similar to those of Part~1.

That the differences in retweet and mention networks are so proportionately similar across both Parts~1 and~2 is notable because the retweet network is not based on direct interactions, while the mention network is. Retweeting a tweet links a retweeter, X, back to the original author, Y, of a tweet, rather than any intermediate account, even if the retweet passed through several accounts on its way between Y and X. It is possible that these datasets were sufficiently constrained both in size and timespan and focus of the participants (by which we mean they engaged in the discussion by following the \texttt{\#qanda} hashtag), that there was little opportunity to build up chains of retweets.

Next we look at two major categories of network analysis: \emph{indexing}, for the computation of node-level properties, such as centrality; and \emph{grouping}, for the computation of specific groups of nodes, such as clustering.

\subsubsection{Comparison of centralities}

Centrality measures can tell us about the influence an individual has over their neighbourhood, though the timing of interactions should ideally be taken into account to get a better understanding of their dynamic aspects (e.g., \citealt{falzon2018embedding}). 
If networks are constructed from partial data, network-level metrics (e.g., radius, shortest paths, cluster detection) and neighbourhood-aware measures (e.g., eigenvector and Katz centrality) may vary and not be meaningful~\citep{HolzmannAK2018}.

\begin{figure}[!ht]
    \centering
    \begin{minipage}[t]{0.99\columnwidth}
        \subfloat[Q\&A Part 1.\label{fig:qanda1_centrality_ranking_comparisons}]{
            \includegraphics[width=\columnwidth]{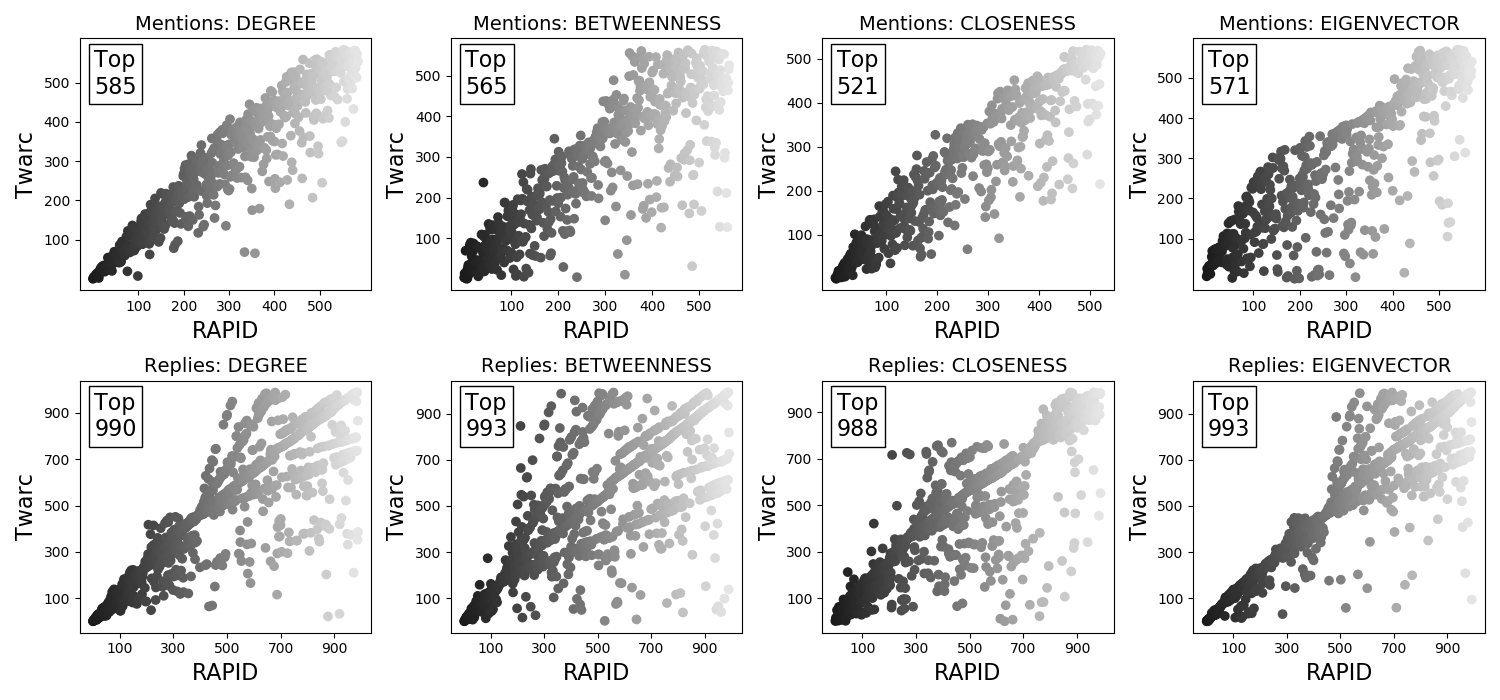}
        }
        \hfill
        \subfloat[Q\&A Part 2. 
        \label{fig:qanda2_centrality_ranking_comparisons}]{
            \includegraphics[width=\columnwidth]{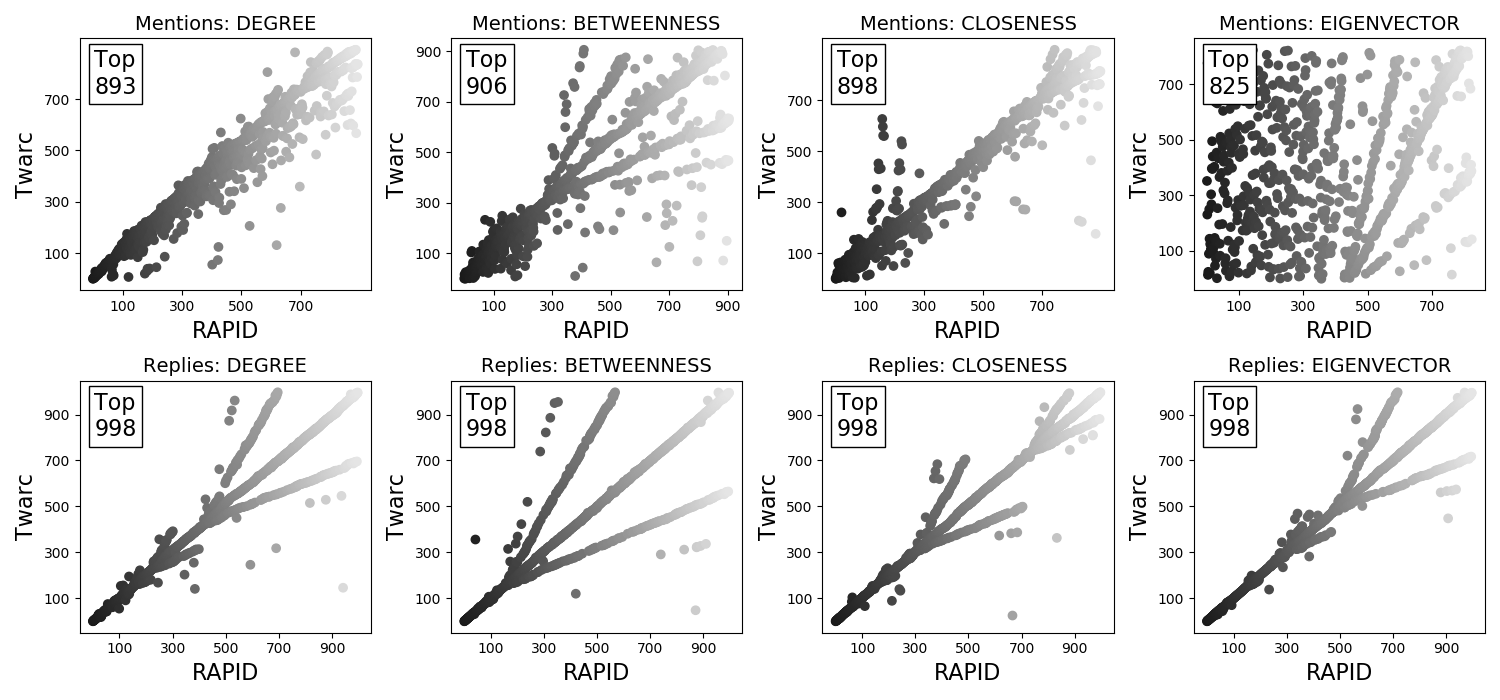}
        }
    \end{minipage}
    \caption{Centrality ranking comparison scatter plots of the mention and reply networks built from the Q\&A Parts 1 and 2 datasets. In each plot, each point represents a node's ranking in the RAPID and Twarc lists of centralities (common nodes amongst the top $1,000$ of each list). The number of nodes appearing in both lists is inset. Point darkness indicates rank on the $x$ axis (darker = higher).}
    \label{fig:qanda_centrality_ranking_comparisons}
\end{figure}

We compare centralities of corresponding networks using scatter plots of node rankings, as per Section~\ref{sec:analyses} (Figure~\ref{fig:qanda_centrality_ranking_comparisons}). 
The symmetrical structures come from corresponding shifts in order: if an item appears higher in one list, then it displaces another, leading to the evident fork-like patterns. 
There is considerable variation in most centrality rankings for both mention and reply networks in Part 1 (Figure \ref{fig:qanda1_centrality_ranking_comparisons}) but much less in Part 2 (Figure \ref{fig:qanda2_centrality_ranking_comparisons}), 
apart from the ranking of eigenvector centralities for the mention networks, which lacks almost any alignment between the RAPID and Twarc node rankings, despite the high number of common nodes ($825$). This implies that the neighbourhoods of nodes differ between the Twarc and RAPID mention networks, but the top ranked nodes are similar though their orders differ greatly. 
Furthermore, the relatively few common nodes in Part 1's Twarc mention networks ($521$ to $585$) and greater edge count (Tables~\ref{tab:qanda1_graph_stats}) could indicate that the extra edges significantly affect the node rankings. 
However, Part 2's Twarc mentions networks also had many more edges, but many more nodes in common (approximately $900$).
Thus it must have been how the mentions were distributed in the datasets that differed, rather than simply their number.
It is not clear that Part 1's four hour duration (\emph{cf.} Part 2's $15$ hours) explains this. 
Instead, if we look at the $11,480$ tweets unique to Twarc in Part~1 (\emph{cf.} fewer than $4,000$ are unique to Twarc in Part 2, Table~\ref{tab:qanda_collection_stats}), 
only $622$ are replies, whereas $6,915$ include mentions. 
There are also $34\%$ more unique accounts in the Part~1 Twarc dataset, but only $19\%$ more in the Part 2 Twarc dataset (Table~\ref{tab:qanda_collection_stats}). 
Each mention refers to one of these accounts and forms an extra edge in the mention network, thus altering the network's structure and the centrality values of many of its nodes; this is likely where the variation in rankings originates.

\begin{figure}[!ht]
    \centering
    \begin{minipage}[t]{\columnwidth}
        \subfloat[Q\&A Part 1.\label{fig:qanda1_centrality_ranking_comparisons_tau_rho}]{
            \includegraphics[width=0.99\textwidth]{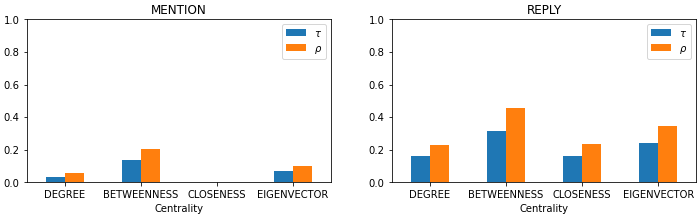}
        }
        \\ 
        \subfloat[Q\&A Part 2.\label{fig:qanda2_centrality_ranking_comparisons_tau_rho}]{
            \includegraphics[width=0.99\textwidth]{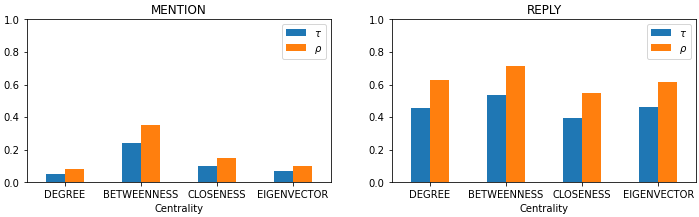}
        }
    \end{minipage}
    \caption{Centrality ranking comparisons using Kendall $\tau$ and Spearman's $\rho$ coefficients
    for corresponding mention and reply networks made from the Q\&A Parts 1 and 2 datasets.}
    \label{fig:qanda_centrality_ranking_comparisons_tau_rho}
\end{figure}


The Kendall $\tau$ and Spearman's $\rho$ coefficients were calculated comparing the corresponding lists of nodes, each pair ranked by one of the four centrality measures 
(Figure~\ref{fig:qanda_centrality_ranking_comparisons_tau_rho}). 
Although somewhat proportional, it is notable how different the coefficient values are, especially in Part~2. 
While Twarc produced more tweets than RAPID (Table~\ref{tab:qanda_collection_stats})
, and more unique accounts, the corresponding mention and reply node counts are not significantly higher (Tables~\ref{tab:qanda1_graph_stats} and~\ref{tab:qanda2_graph_stats}). 
In fact, the node counts in the Part~1 reply networks are correspondingly lower than in the Part~2 reply networks, 
even though both Part~2 datasets were smaller.
Edge counts in the mention networks were very different (Twarc had many more) but were quite similar in the reply networks.

The biggest variation was in the mention networks from Part 1 (Figure~\ref{fig:qanda1_centrality_ranking_comparisons} and Table~\ref{tab:qanda1_graph_stats}), 
due to the large number of extra mentions from Twarc. It is notable that the Kendall's $\tau$ was low for all mention networks (Figure~\ref{fig:qanda_centrality_ranking_comparisons_tau_rho}), especially for degree and closeness centrality. 
It is worth noting the minor differences in the degree and immediate neighbours of nodes impacts degree and closeness centralities significantly, and, correspondingly, their relative rankings. 
In contrast, rankings for betweenness and eigenvector centrality, which rely more on global network structure, remained relatively stable.

\subsubsection{Comparison of clusters}

\begin{figure}[!ht]
    \begin{minipage}[t]{\columnwidth}
        \subfloat[Q\&A Part 1.\label{fig:qanda1_top20_clusters}]{
            \includegraphics[width=0.47\columnwidth]{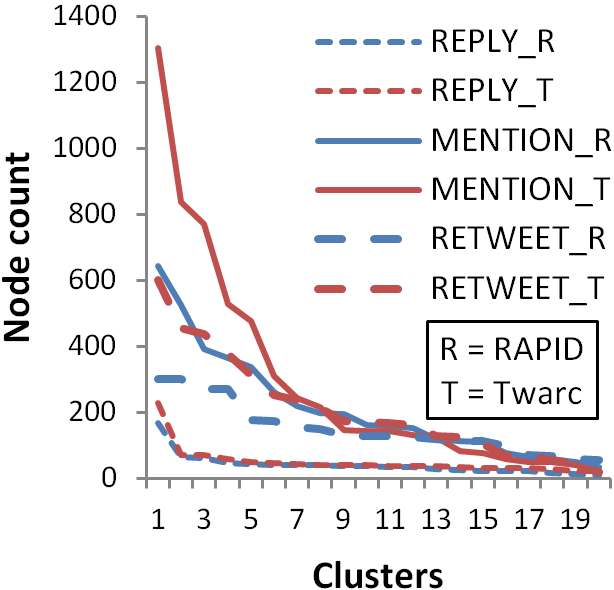}
        }
        \hfill
        \subfloat[Q\&A Part 2.\label{fig:qanda2_top20_clusters}]{
            \includegraphics[width=0.47\columnwidth]{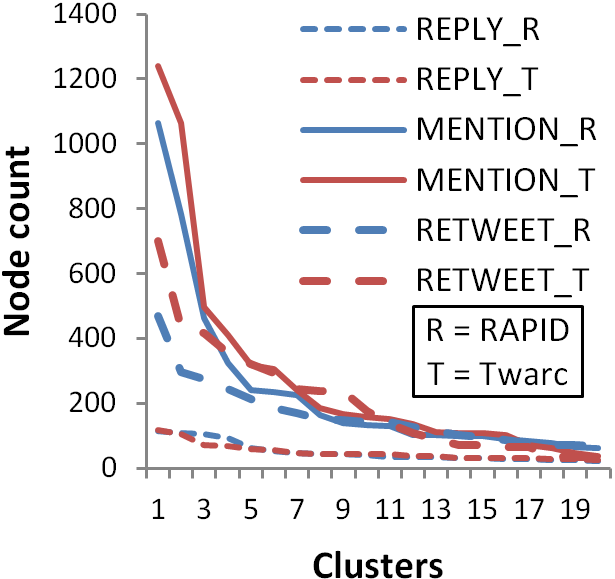}
        }
    \end{minipage}
    \caption{The largest retweet, mention and reply clusters built from the Q\&A Parts 1 and 2 datasets.}
    \label{fig:qanda_top20_clusters}
\end{figure}

We finally compare the networks via largest clusters 
(Figure~\ref{fig:qanda_top20_clusters}).  
The reply network clusters are relatively similar, 
and the largest mention and reply clusters differ the most.
The ARI scores (Table~\ref{tab:qanda_ari_scores}) 
confirm that the reply clusters were most similar for Parts~1 and~2 ($0.738$ and $0.756$, respectively), possibly due to the small size of the reply networks. The mention and retweet clusters for Part~2 were more similar than those of Part~1 ($0.437$ and $0.468$ compared to $0.320$ and $0.350$), possibly due to the longer collection period. In Part~1, there is a chance the networks are different due to RAPID's expansion strategy. 
Changes to filter keywords may have collected  
posts of other vocal accounts not using the original keywords, at the cost of the posts 
which did.



\begin{table}[ht]
    \centering\small
    \begin{tabular}{@{}lrrr@{}}
        \toprule
                     & RETWEET & MENTION & REPLY \\
        \midrule
        Q\&A Part 1  & 0.320   &  0.350  & 0.738 \\
        Q\&A Part 2  & 0.437   &  0.468  & 0.756 \\
        \bottomrule
    \end{tabular}
    \caption{Adjusted Rand index scores for the clusters found in the corresponding retweet, mention and reply networks built from the Q\&A Parts 1 and 2 datasets.}
    \label{tab:qanda_ari_scores}
\end{table}

\subsubsection{Summary of findings}

Overall, Twarc and RAPID provided very different views into the Twitter activity surrounding the Q\&A episode in question, both on the evening of and the day after. 
This includes variations in basic collection statistics, network statistics for retweet, mention and reply networks built from the collected data, centrality measures of the nodes in the networks and comparison of detected clusters. 
The extra tweets collected by the Twarc collections appear to have resulted in greater numbers of connections internal to the largest components, which may have implications for the analysis of influence, as reachability correspondingly increases. Deeper study of reply content is required to inform patterns of information gathering.

We are left with the open question of how reliable social media can be as a data source, if conducting simultaneous collection activities with the same query criteria can provide such different networks. 
Is the variation due to the platform providing a random sample of the overall data 
or an effect of the tool being used?

We next considered a more tightly controlled comparison of Twarc and RAPID, disabling RAPID's expansion strategies so that the tools performed as similarly as possible.

\subsection{Case Study 2: A weekend of AFL without topic tracking} \label{sec:afl}



\begin{figure}[!ht]
    \centering
    \includegraphics[width=\columnwidth]{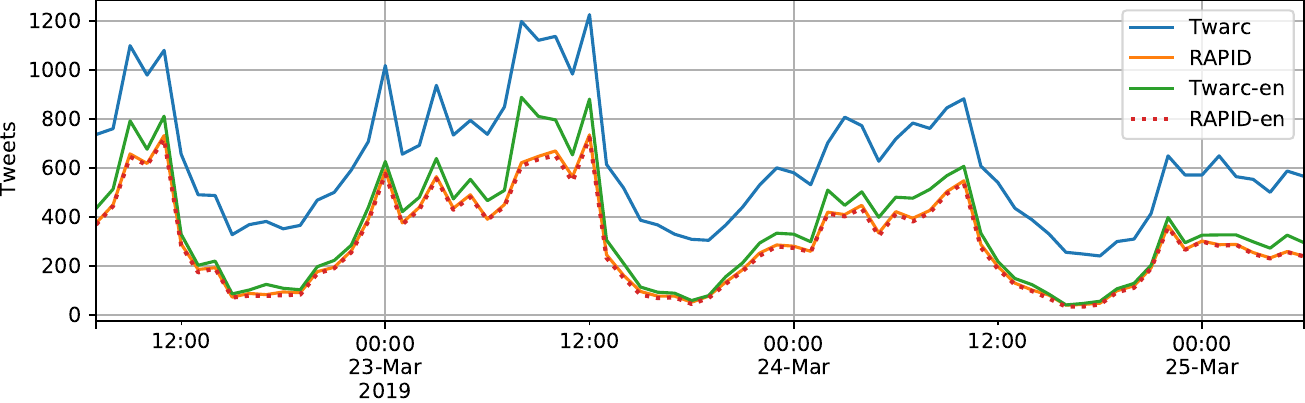}
    \caption{Twitter activity in the AFL1 dataset over time (in 60 minute blocks).}
    \label{fig:afl1_timeline}
\end{figure}

RAPID's topic tracking feature broadens the scope of of the collection at the cost of strictly matching tweets, resulting in distinctly different corresponding corpora. 
Although the rankings of the most central nodes in networks built from the corpora appear relatively stable, the question remains of why the corpora were so different in size. In this section, we discuss a case study in which we disabled RAPID's topic tracking feature, expecting the resulting corresponding corpora to increase in similarity, especially over a longer period collection. Figure~\ref{fig:afl1_timeline} indicates that again, initially at least, it appeared that Twarc and RAPID produced very different, but proportional over time, datasets. Constraining the datasets to only those tweets with a ``lang'' property of ``en'' or ``und'' resulted in much more similar datasets.

\begin{table}[t]
    \centering
    \caption{Summary dataset statistics of the AFL1 collection.}
    \label{tab:afl1_collection_stats}
    \resizebox{\columnwidth}{!}{%
        \begin{tabular}{@{}l|r|rr|rr|r|rr@{}}
        \toprule
        Dataset & \multicolumn{1}{c}{All} & \multicolumn{2}{c}{Unique} & \multicolumn{2}{c}{Retweets} & \multicolumn{1}{c}{All} & \multicolumn{2}{c}{Unique} \\ 
                & \multicolumn{1}{c}{Tweets} & \multicolumn{2}{c}{Tweets} & \multicolumn{2}{c}{} & \multicolumn{1}{c}{Accounts} & \multicolumn{2}{c}{Accounts} \\
        \midrule
        Twarc    & 44,461  & 22,731 & (51.1\%) & 11,482   & (25.8\%) & 16,821   & 5,274    & (31.4\%) \\
        RAPID    & 21,799  & 69     & (0.3\%)  & 7,047    & (32.3\%) & 11,573   & 26       & (0.2\%)  \\
        \midrule
        Twarc-en & 25,231  & 4,065  & (16.1\%) & 8,531    & (33.8\%) & 12,399   & 1,187    & (9.6\%)  \\
        RAPID-en & 21,235  & 69     & (0.3\%)  & 6,849    & (32.3\%) & 11,238   & 26       & (0.2\%)  \\
        \bottomrule
        \end{tabular}%
    }
\end{table}

\subsubsection{Comparison of collection statistics}

\begin{figure}[th!]
    \centering
    \includegraphics[width=0.99\columnwidth]{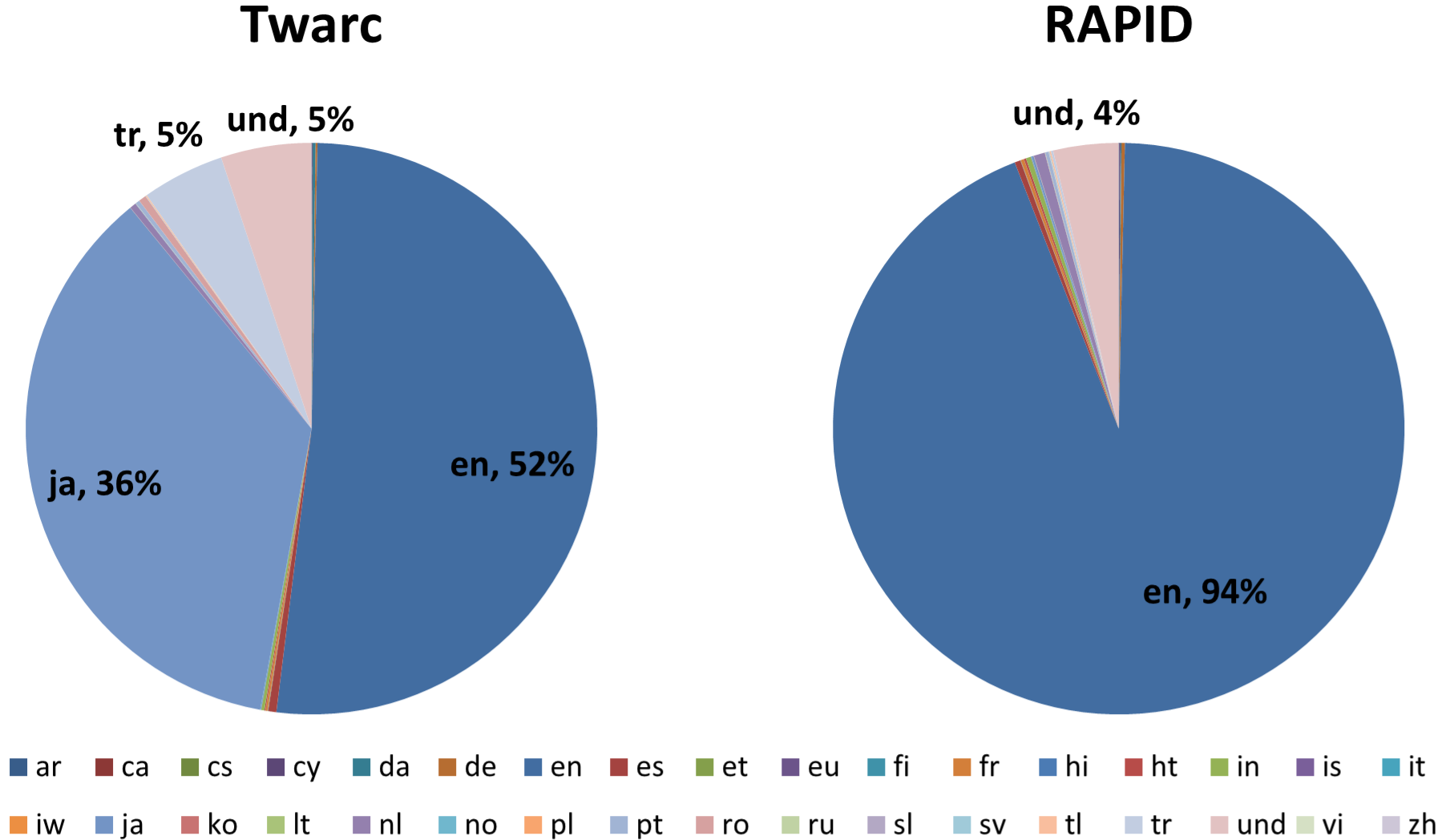}
    \caption{Language distributions of the RAPID and Twarc collections using the filter term ``afl''.}
    \label{fig:afl_tweet_langs}
\end{figure}

\begin{figure}[ht!]
    \centering
    \includegraphics[width=0.75\columnwidth]{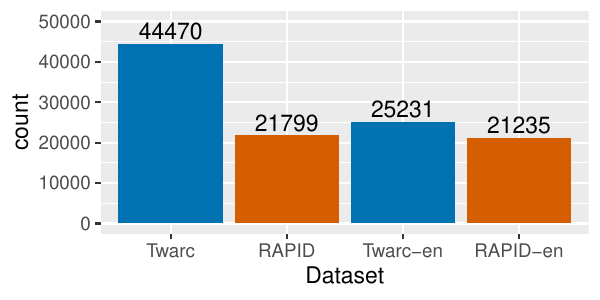}
    \caption{Tweet counts from the RAPID and Twarc collections using the filter term ``afl''.}
    \label{fig:afl_tweet_counts}
\end{figure}

We conducted two parallel collections under the term ``afl'' over a three day period in March 2019 using RAPID without topic tracking and Twarc. This collection is labelled ``AFL1'' in Tables~\ref{tab:collection_conditions} and~\ref{tab:all_collection_stats} and further detail is offered in Table~\ref{tab:afl1_collection_stats}. The datasets obtained appear to be dramatically different: RAPID collected just shy of 22k tweets while Twarc found approximately twice that number with around 45k tweets, with 21,730 in common.  Interestingly, as can be seen in Figure~\ref{fig:afl1_timeline}, the extra tweets appear to occur relatively evenly and consistently over time, rather than spiking. On closer inspection, it became apparent that the balance in languages was different, with 36\% of the Twarc collection having \texttt{lang} property of `jp' (Japanese) and 52\% `en' (English), while RAPID consisted of 94\% English tweets (Figure~\ref{fig:afl_tweet_langs}). When both collections were trimmed to tweets with a \texttt{lang} property of `en' or `und' (undefined), they reduced to $25,231$ tweets (Twarc) and $21,235$ tweets (RAPID), with 21,166 in common, which still leaves more than 4,000 tweets specific to Twarc (Figure~\ref{fig:afl_tweet_counts}). Twitter populates this field based on a variety of factors including the account's preferred language but also on built-in language detection facilities, with varying degrees of success. 
The ``AFL1'' dataset, reduced to only posts with a \texttt{lang} property of `en' or `und' is referred to as ``AFL1-en'' henceforth.

As previously mentioned, RAPID does not retain tweets which do not contain filter terms in text-related portions of the tweets. 
In the Twarc collection, the term `afl' appeared in the domain of a website that many of the Japanese tweets referred to, belonging to an online marketplace. These tweets were dropped by RAPID and did not appear in the final collection. 

Only $69$ tweets were unique to the RAPID AFL1-en dataset, and they appear to be AFL-related sports discussions. The $4,065$ tweets unique to the Twarc dataset comprise $2,595$ English tweets and $1,470$ with ``und'' for the \texttt{lang} value. This field is populated by Twitter based on language detection algorithms. When a language cannot be detected, such as when there is not sufficient free text to analyse, the value ``und'' is used. Inspection of the tweets indicates the reason for this: the undefined tweets include $884$ retweets, $1,366$ tweets with URLs, $116$ with hashtags, and $916$ with mentions. Of the ``und'' tweets with URLs, the vast majority ($1,188$) refer to a Japanese online electronics marketplace ($771$) and a Japanese online media platform ($417$). The next largest group refer to $38$ retweets, some of the official @AFL account ($9$), though there are $16$ and $5$ retweets of two accounts that Botometer \citep{botornot2016} scored at $4.2$ and $4.4$ out of $5$, respectively, as bot-like, and both refer to the previously mentioned Japanese electronics marketplace. The top $12$ most used hashtags in the English subset relate to the AFL, while the top $14$ for the ``und'' subset are all Japanese terms, except for ``iphone'' (at number 9). The top term (in Japanese) is the name of the marketplace. 
The English tweets are mostly related to the AFL, though there is considerable obvious content from bot-like accounts, with several accounts posting the same content (offers of live streams of the matches) repeatedly within a short space of time (their messages appear adjacent in the timeline).

\begin{table}[t]
    \centering
    \resizebox{\columnwidth}{!}{%
    \begin{tabular}{@{}lrr@{}}
        \toprule

        Property                           & Twarc-en                         & RAPID-en                         \\
        \midrule
        Tweets                             &                           25,231 &                           21,235 \\
        \midrule
        Accounts                           &                           12,399 &                           11,238 \\
        Retweets                           &                            8,531 &                            6,849 \\
        Quotes                             &                            2,291 &                            1,615 \\
        Replies                            &                            6,185 &                            5,936 \\
        Tweets with hashtags               &                            7,606 &                            6,911 \\
        Tweets with URLs                   &                           10,266 &                            7,345 \\
        Most prolific account              & Account $a_5           $         & Account $a_5                   $ \\
        Tweets by most prolific account    &                              363 &                              362 \\
        Most retweeted tweet               & Tweet $t_5$                      & Tweet $t_5$                      \\
        Most retweeted tweet count         &                              529 &                              529 \\
        Most replied to tweet              & Tweet $t_6$                      & Tweet $t_6$                      \\
        Most replied to tweet count        &                              141 &                              141 \\
        Tweets with mentions               &                           17,467 &                           15,230 \\
        Most mentioned account             & Account $a_6$                    & Account $a_6$                    \\
        Mentions of most mentioned account &                            7,131 &                            7,130 \\
        Hashtag uses                       &                           17,352 &                           15,886 \\
        Unique hashtags                    &                            2,381 &                            2,249 \\
        Most used hashtag                  & \texttt{\#afl}                   & \texttt{\#afl}                   \\
        Most used hashtag count            &                            4,523 &                            4,522 \\
        Next most used hashtag             & \texttt{\#aflpiescats}           & \texttt{\#aflpiescats}           \\
        Uses of next most used hashtag     &                            1,575 &                            1,482 \\
        URL uses                           &                            6,557 &                            3,552 \\
        Unique URLs                        &                            2,843 &                            2,043 \\
        Most used URL                      & \url{http://watchrugby.net/AFL/} & \url{http://watchrugby.net/AFL/} \\
        Uses of most used URL              &                              494 &                              251 \\
        \bottomrule
    \end{tabular}
    } 
    \caption{Statistics of the AFL1-en RAPID and Twarc datasets.}
    \label{tab:afl1-en_dataset_stats}
\end{table}

Once reduced to a relatively comparable state, the ``AFL1-en'' parallel datasets can be examined in more detail. It is understood that the tweets they consist of will differ, given that rate-limiting constraints may have caused each to receive different tweets. The statistics in Table \ref{tab:afl1-en_dataset_stats} bear this out with the Twarc dataset statistics being approximately proportionately larger when compared with the RAPID dataset statistics. 
The author IDs have been anonymised, but the most mentioned account is the official @AFL account, while the most prolific author appears to be automated to some degree, having posted nearly $35,000$ tweets in two years and a Botometer \citep{botornot2016} Complete Automation Probability (CAP\footnote{See \url{https://botometer.osome.iu.edu/faq#which-score}.}) of $68\%$, many seemingly promote the AFL, tennis, and a singer. The most replied to tweet was posted by an Australian NBA\footnote{United States National Basketball Association} player and the most retweeted tweet was of an amusing video of an AFL supporter.

\subsubsection{Comparison of network statistics}

\begin{table}[t]
    \centering
    \resizebox{\columnwidth}{!}{%
    \begin{tabular}{@{}lrrrrrr@{}}
        \toprule
                           & \multicolumn{2}{c}{RETWEET} & \multicolumn{2}{c}{MENTION} & \multicolumn{2}{c}{REPLY} \\ 
                           & RAPID-en    & Twarc-en      & RAPID-en    & Twarc-en      & RAPID-en    & Twarc-en    \\
                           \cmidrule(lr){2-3}            \cmidrule(lr){4-5}            \cmidrule(lr){6-7}
        Number of nodes	   & 5,584	     & 6,430         & 11,566      & 12,525	       & 4,705	     & 4,759       \\
        Number of edges	   & 5,881	     & 6,977         & 22,310      & 23,937	       & 4,928	     & 5,005       \\
        Average degree     & 1.053 	     & 1.085 	     & 1.929	   & 1.911 	       & 1.047 	     & 1.052       \\
        Density	           & 0.000       & 0.000     	 & 0.000       & 0.000         & 0.000       & 0.000       \\
        Mean edge weight   & 1.165       & 1.223     	 & 1.308       & 1.329         & 1.205       & 1.236       \\
        Components   	   & 494	     & 536	         & 666	       & 713	       & 791	     & 798         \\
        Largest component  & 3,946       & 4,233         & 9,416       & 9,789	       & 2,951	     & 3,017       \\
        - Diameter         & 16          & 16            & 17          & 17            & 16          & 16          \\
        Clusters   	       & 544	     & 579	         & 736	       & 781	       & 861	     & 863         \\
        Largest cluster	   & 659	     & 753	         & 2,177       & 2,125         & 851	     & 844         \\
        Reciprocity	       & 0.004       & 0.005     	 & 0.057       & 0.055         & 0.139       & 0.131       \\
        Transitivity	   & 0.035       & 0.038     	 & 0.143       & 0.152         & 0.039       & 0.034       \\
        Maximum k-core     & 4	         & 4	         & 9           & 11	           & 4	         & 4           \\
        \bottomrule
    \end{tabular}
    } 
    \caption{Comparative statistics for networks generated from the RAPID and Twarc datasets for the ALF1-en collection.}
    \label{tab:afl1-en_graph_stats}
\end{table}


\begin{figure}[!ht]
    \centering
    \includegraphics[width=0.99\columnwidth]{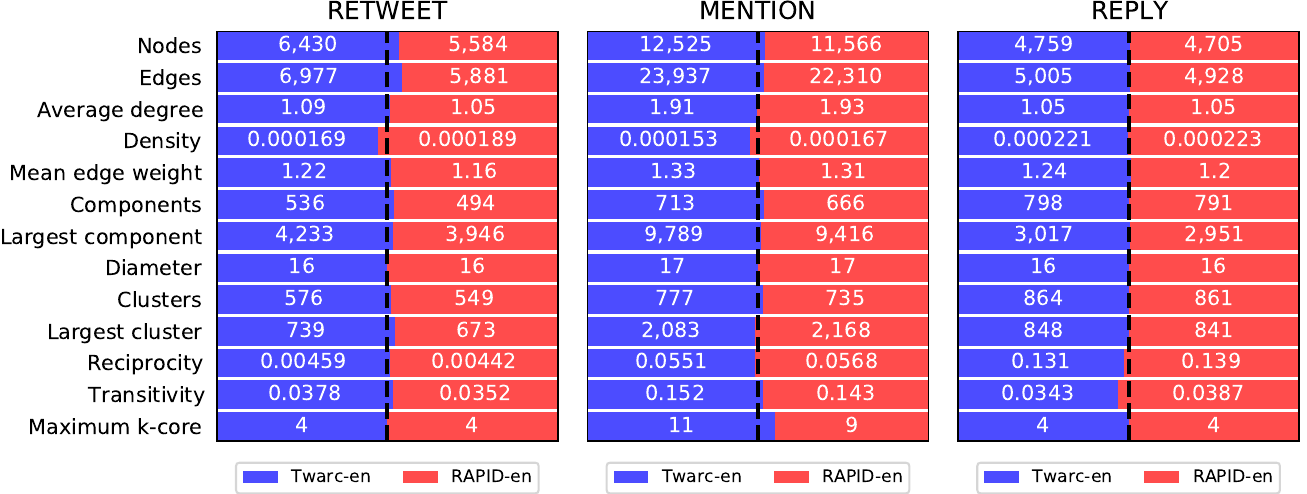}
    \caption{The proportional balance betweet Twarc and RAPID statistics of the retweet, mention and reply networks built from the AFL1-en datasets.}
    \label{fig:afl1_graph_stats_barh}
\end{figure}

The network statistics Table~\ref{tab:afl1-en_graph_stats} indicate that the networks were much more similar than in the Q\&A case study, though there are still notable differences. The largest components of the retweet, mention and reply networks are, at most, 15\% larger by node count, and the largest components are correspondingly similar, though their diameters and densities indicate they are much more sparse than the corresponding Q\&A ones, with corresponding implications for the opportunity to influence. In contrast, in sporting discussions, there is less motivation to attempt to convert fellow sports fans to cheer for one's team than there is in a political discussion. 
Certainly in this study, politics has tended to generate more discussion than sports in general, and the nature of the discussions is also different. 
The reciprocity values here are much higher than in the Q\&A case study, implying the presence of more communication among the communities that do exist. Another difference that lends weight to this interpretation is the average degree of nodes in the networks. In the Q\&A retweet and mention networks, the average degrees were around 2-2.5 and 3, respectively, implying some repetition in connectivity, whereas in the sporting discussing the average degrees are around 1 and 2, respectively, implying much less continued interaction. As indicated in Figure~\ref{fig:afl1_graph_stats_barh}, the degree values of the Twarc and RAPID networks are highly similar.

The number of tweets and accounts in the AFL1-en datasets (Table~\ref{tab:afl1-en_dataset_stats}), coupled with the number of nodes and edges in the derived mention and reply networks (Table~\ref{tab:afl1-en_graph_stats}), 
indicate that although the AFL1-en collections differed by nearly $4,000$ tweets, the number of accounts was not significantly different (approximately $10\%$ more in Twarc) with a corresponding increase in nodes and edges in the mention network (8.3\% and 7.3\%, respectively) but only $54$ and $77$ (1.1\% and 1.6\%, respectively) more in the reply network.


\subsubsection{Comparison of centralities}
\begin{figure}[t!]
    \centering
    \includegraphics[width=\columnwidth]{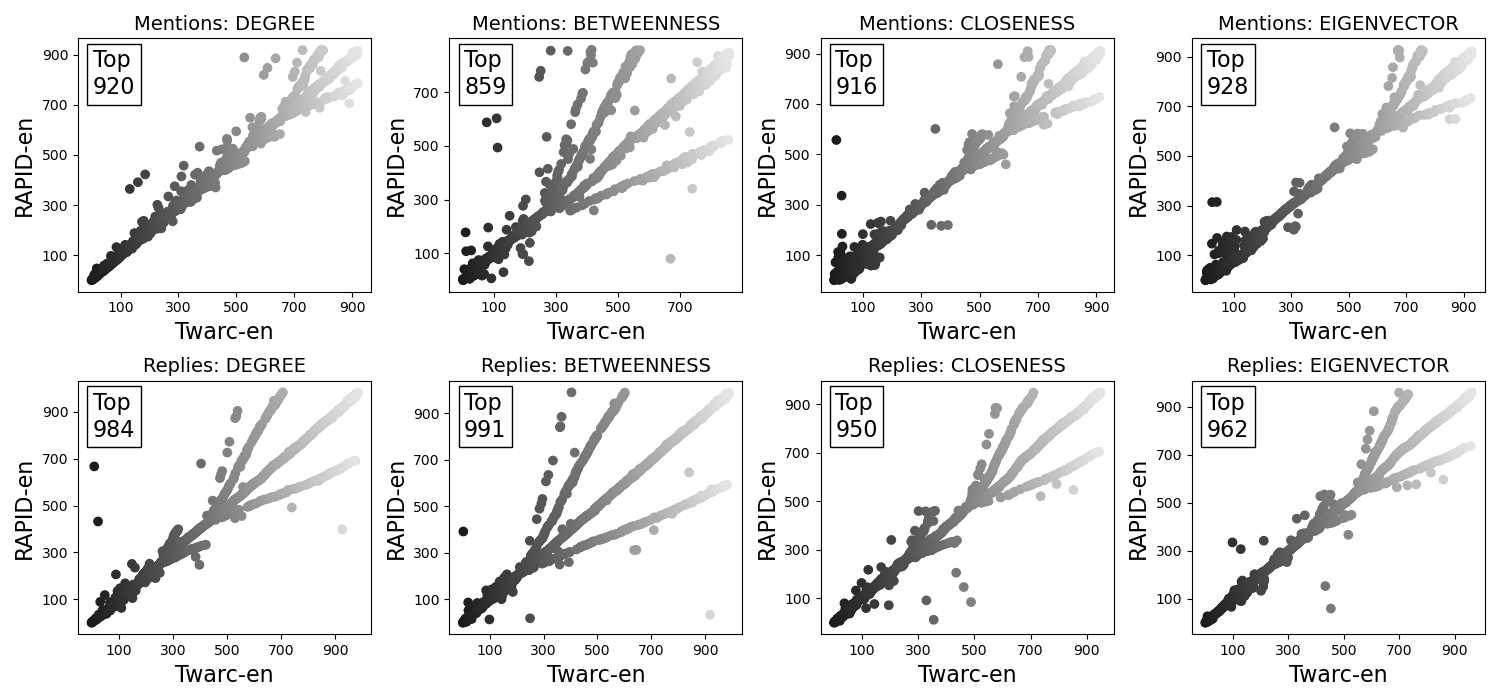}
    \caption{Centrality ranking comparison scatter plots of the mention and reply networks built from the AFL1-en datasets. In each plot, each point represents a node’s ranking in the RAPID-en and Twarc-en lists of centralities (common nodes amongst the top $1,000$ of each list). The number of nodes appearing in both lists is inset. Point darkness indicates rank on the $x$ axis (darker = higher).}
    \label{fig:afl1-en_und_centrality_ranking_comparisons_scatterplots}
\end{figure}

\begin{figure}[t!]
    \centering
    \includegraphics[width=0.99\columnwidth]{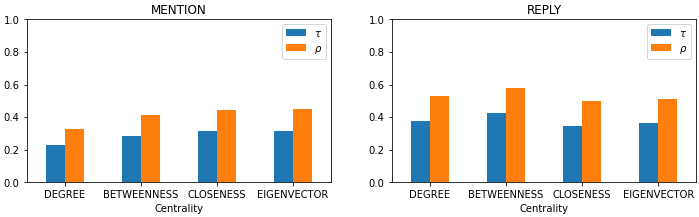}
    \caption{Centrality ranking comparisons from the RAPID and Twarc datasets of the AFL1-en collection using Kendall Tau scores and Spearman's Coefficients.}
    \label{fig:afl1-en_und-centrality_ranking_comparisons_tau_rho}
\end{figure}

Considering the similarity of interaction networks constructed from the respective AFL1-en datasets, we compare the relative ranking of the the top network nodes by various centrality values (with an upper bound of $1,000$ nodes). Figure~\ref{fig:afl1-en_und_centrality_ranking_comparisons_scatterplots} shows scatterplots of the relative rankings of nodes common to corresponding networks, and Figure~\ref{fig:afl1-en_und-centrality_ranking_comparisons_tau_rho} plots the Kendall Tau and Spearman's coefficients of the corresponding relative rankings. As with the Q\&A collection, the centrality of nodes in the reply networks show more similarity than those in the mention networks, which is likely due to their relative size; Table~\ref{tab:afl1-en_graph_stats} indicates a significant discrepancy in the reply and mention network sizes and average degree. 
Closeness is notably low in similarity, though the high component count would account for that. 
It is apparent the the most central nodes in both network types mostly maintain their ordering for the first several hundred nodes, but all begin to diverge at some point. A few isolated nodes change their ranking significantly, such as those in the top left of the mentions betweenness and closeness plots, degrading their rankings (appearing above the diagonal), and those in the reply closeness and eigenvector plots, improving their rankings (appearing below the diagonal), but the majority diverge in a trident pattern, implying lower ranked nodes improve their rankings swapping out higher ranked nodes at progressively greater distances. The reason for the consistency is unclear.  
Minor variations would ensure that nodes' centrality values varied and thus their rankings could easily vary significantly, especially due to the high number of components. The high K-core values for the mention networks are likely to explain the high betweenness and eigenvector centrality values, as the highest ranked of these will reside in the larger components, which will have the greater likelihood of being similar across the networks.

\subsubsection{Comparison of clusters}

Comparing the clusters detected with the Louvain method \citep{blondel2008louvain} in the retweet, mention and reply networks results in ARI values in Table~\ref{tab:afl1-en_ari_scores}. This implies that although the networks consisted of many components, the clusters they formed were highly similar for retweet and reply networks, and only slightly less so for the mention networks, despite the fact that the Twarc mention network included more than $2,000$ more mention edges.

\begin{table}[ht]
    \centering\scriptsize
    \begin{tabular}{@{}rrr@{}}
        \toprule
        RETWEET & MENTION & REPLY \\
        \midrule
        0.818   &  0.675  & 0.853 \\
        \bottomrule
    \end{tabular}
    \caption{Adjusted Rand index scores for the clusters found in the networks built from the RAPID and Twarc datasets for the AFL1-en collection.}
    \label{tab:afl1-en_ari_scores}
\end{table}

\subsubsection{Summary of findings}

This case study makes it clear that the tool used for collection can have a significant effect on the data collected and the resulting analytic results. 
It was serendipitous that the filter term chosen was ``afl'', because a more specific term or set of terms is unlikely to have captured the non-English content that Twarc did. This highlighted the fact that RAPID was post-processing and filtering the tweets it collected,
and raises general questions for social media data collection:
Do other collection tools, especially commercial ones, do this post-processing too, as a ``convenience'' or ``value-add'' to their users? 
Do they make it clear if and when they do? 
The validity of evidence-based conclusions rests on these details. 
Even when both datasets were filtered to ensure some degree of consistency, there remained large differences in the networks constructed from them.
Minor differences in datasets may result in amplified differences in analyses.

A further, even more fundamental, question remained after this case study, which is addressed by the next subsection: \emph{Does the same tool provide the same data over two simultaneous collections with identical filter terms?}

\subsection{Case Study 3: Tracking AFL Twitter activity with RAPID}

\begin{table}[t]
    \centering
    \caption{Summary dataset statistics of the AFL2 collection.}
    \label{tab:afl2_collection_stats}
    \resizebox{\columnwidth}{!}{%
        \begin{tabular}{@{}l|r|rr|rr|r|rr@{}}
        \toprule
        Dataset & \multicolumn{1}{c}{All} & \multicolumn{2}{c}{Unique} & \multicolumn{2}{c}{Retweets} & \multicolumn{1}{c}{All} & \multicolumn{2}{c}{Unique} \\ 
                & \multicolumn{1}{c}{Tweets} & \multicolumn{2}{c}{Tweets} & \multicolumn{2}{c}{} & \multicolumn{1}{c}{Accounts} & \multicolumn{2}{c}{Accounts} \\
        \midrule
        RAPID1  & 30,103 & -      & (0.0\%) & 9,215    & (30.6\%) & 14,231   & -        & (0.0\%) \\
        RAPID2  & 30,115 & 12     & (0.0\%) & 9,215    & (30.6\%) & 14,232   & 1        & (0.0\%) \\
        \bottomrule
        \end{tabular}%
    }
\end{table}

\begin{figure}[!ht]
    \centering
    \includegraphics[width=\columnwidth]{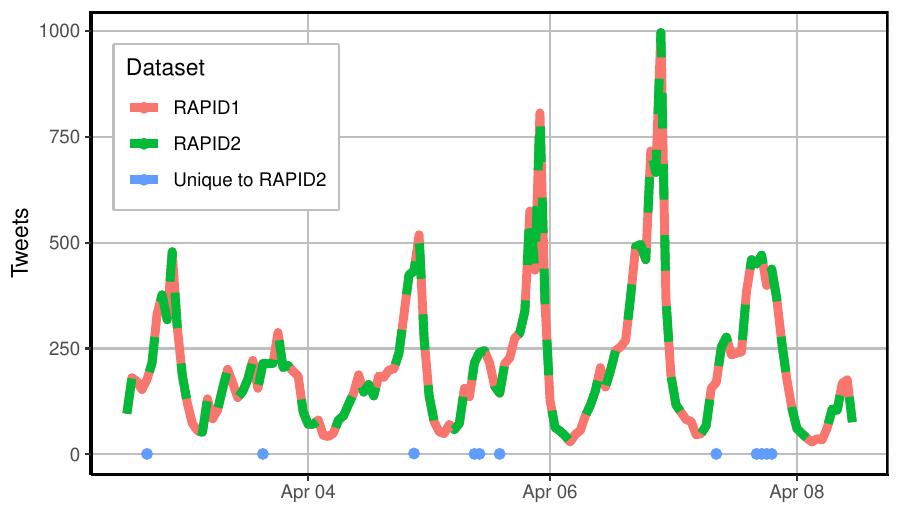}
    \caption{Twitter activity in the AFL2 dataset over time (in 60 minute blocks). Thick and dashed lines are used here to highlight how the timeseries overlap almost exactly. The timestamps of tweets unique to RAPID2 are shown as blue points.}
    \label{fig:afl2_timeline}
\end{figure}

Given it appeared that different collection tools could produce different results using the same inputs, the question of whether APIs are delivering consistent content for all clients remained. 
A second collection (``AFL2'' in Table \ref{tab:qanda_collection_stats}) was initiated over a longer period (six days) using the same filter term and tool (RAPID), but with different API credentials. 
One set of credentials belonged to a relatively new and unused account (created in 2018 having posted only 3 tweets) and the other to a well-established account (created in 2009 and having posted 17k tweets). 
This resulted in two highly similar, but not quite identical, datasets, with sizes $30,103$ and $30,115$ tweets; their timeline is shown in Figure~\ref{fig:afl2_timeline}. 
The first dataset was a proper subset of the second, so the difference of $12$ posts can be regarded as due to noise or minor differences in timing. 
A brief examination revealed these extra tweets (shown in blue in the Figure) were all about AFL or other sports in Australia, and their timing appeared random.  
Further confirmation of the similarity between datasets can be seen in Table~\ref{tab:afl1_vs_afl2} where the most prolific account, most retweeted tweet, most replied to tweet and most mentioned account details are all identical. Again, the most mentioned account is the official @AFL account.  

\begin{table}[t]
    \centering
    \resizebox{\columnwidth}{!}{%
    \begin{tabular}{@{}lrr@{}}
        \toprule
        Property                             & RAPID1                           & RAPID2                           \\
        \midrule
        Tweet count                          & 30,103                           & 30,115                           \\
        \midrule
        Retweet count                        & 9,215                            & 9,215                            \\
        Account count                        & 14,231                           & 14,232                           \\
        Quote count                          & 2,340                            & 2,341                            \\
        Reply count                          & 9,229                            & 9,229                            \\
        Tweets with hashtags                 & 8,623                            & 8,627                            \\
        Tweets with URLs                     & 11,467                           & 11,474                           \\
        Most prolific account                & Account $a_7$                    & Account $a_7$                    \\
        Most prolific account tweet count    & 612                              & 612                              \\
        Most retweeted tweet                 & Tweet $t_7$                      & Tweet $t_7$                      \\
        Most retweeted tweet count           & 269                              & 269                              \\
        Most replied to tweet                & Tweet $t_8$                      & Tweet $t_8$                      \\
        Most replied to tweet count          & 206                              & 206                              \\
        Tweets with mentions                 & 22,083                           & 22,083                           \\
        Most mentioned account               & Account $a_6$                    & Account $a_6$                    \\
        Most mentioned account count         & 10,468                           & 10,468                           \\
        Hashtag uses                         & 20,136                           & 20,140                           \\
        Unique hashtags                      & 3,337                            & 3,337                            \\
        Most used hashtag                    & \texttt{\#afl}                   & \texttt{\#afl}                   \\
        Most used hashtag count              & 5,096                            & 5,096                            \\
        Next most used hashtag	             & \texttt{\#afldeesdons}           & \texttt{\#afldeesdons}           \\
        Uses of next most used hashtag       &759                               & 759                              \\
        URL uses                             & 5,702                            & 5,709                            \\
        Unique URLs                          & 3,580                            & 3,587                            \\
        Most used URL                        & \url{http://watchrugby.net/AFL/} & \url{http://watchrugby.net/AFL/} \\
        Most used URL count                  & 341                              & 341                              \\
        \bottomrule
    \end{tabular}
    } 
    \caption{Statistics of two parallel datasets collected using RAPID with the filter term ``afl'' over a six day period with different API credentials.}
    \label{tab:afl1_vs_afl2}
\end{table}

\subsubsection{Comparison of network statistics}

\begin{table}[b]
    \centering\scriptsize
    \begin{tabular}{@{}lrrrrrr@{}}
        \toprule
                &        &          & \multicolumn{2}{c}{MENTION} & \multicolumn{2}{c}{REPLY} \\
                                  \cmidrule(lr){4-5}            \cmidrule(lr){6-7}
        Dataset & Tweets & Accounts & Nodes  & Edges              & Nodes & Edges             \\
        \midrule
        RAPID1  & 30,103 &   14,231 & 15,323 & 31,859             & 6,778 & 7,655             \\  
        RAPID2  & 30,115 &   14,232 & 15,323 & 31,859             & 6,778 & 7,655             \\  
        \bottomrule
    \end{tabular}
    \caption{Selected comparative statistics for networks generated from the two RAPID datasets for the AFL2 collection.}
    \label{tab:afl2_selected_graph_stats}
\end{table}

\begin{figure}[!ht]
    \centering
    \includegraphics[width=0.99\columnwidth]{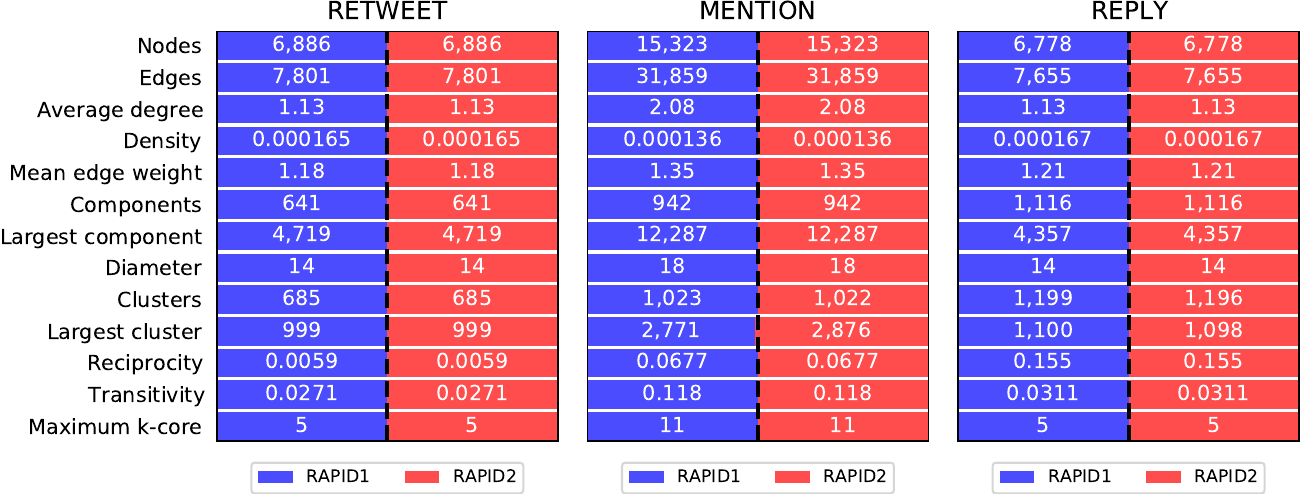}
    \caption{The proportional balance between the RAPID1 and RAPID2 statistics of the retweet, mention and reply networks built from the AFL2 datasets.}
    \label{fig:afl2_graph_stats_barh}
\end{figure}

Due to the similarity of the datasets, the retweet, mention and reply networks generated from them were almost identical, and only a summary of the structures is provided in Table~\ref{tab:afl2_selected_graph_stats}. Details of the networks are provided in Figure~\ref{fig:afl2_graph_stats_barh}, which show that the only differences occur in the detected clusters. In particular, the largest cluster detected in the RAPID2 mention network is around $3\%$ larger than the corresponding cluster from the RAPID1 mention network. 
This is likely due to an element of randomness used in the Louvain algorithm~\citep{blondel2008louvain}.


\begin{figure}[!ht]
    \centering
    \includegraphics[width=0.99\columnwidth]{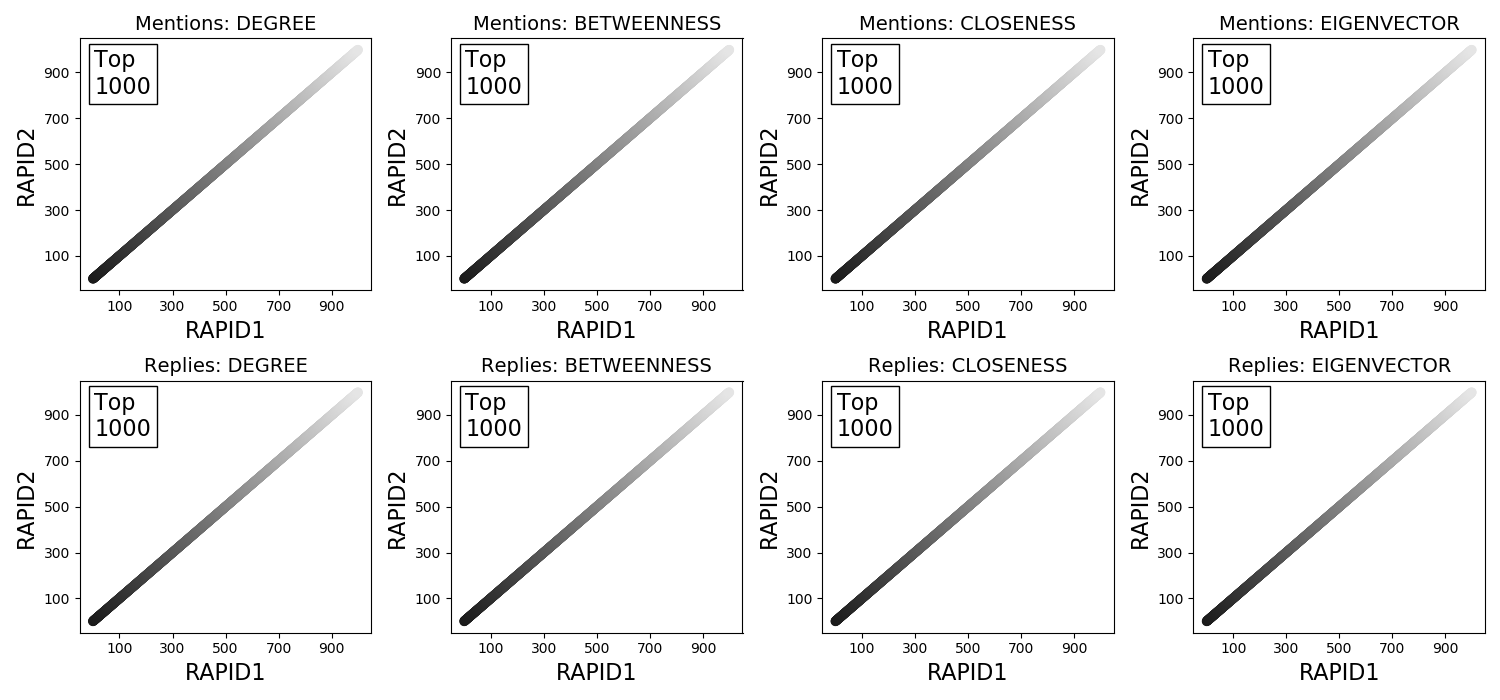}
    \caption{Centrality  ranking  comparison  scatter  plots  of  the  mention  and  reply  networks built from the AFL2 datasets. In each plot, each point represents a node’s ranking in the RAPID1 and RAPID2 lists of centralities (common nodes amongst the top $1,000$ of each list). The number of nodes appearing in both lists is inset. Point darkness indicates rank on the $x$ axis (darker = higher).}
    \label{fig:afl2_centrality_scatterplots}
\end{figure}

\begin{figure}[t!]
    \centering
    \includegraphics[width=0.99\columnwidth]{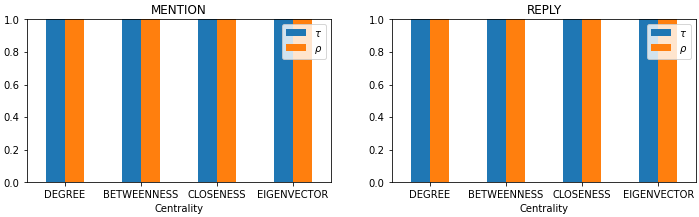}
    \caption{Centrality ranking comparisons from the two RAPID datasets of the AFL2 collection using Kendall Tau scores and Spearman's Coefficients.}
    \label{fig:afl2-centrality_ranking_comparisons_tau_rho}
\end{figure}

\subsubsection{Comparison of centralities and clusters}
The similarity of the networks based on their statistics is further confirmed by a comparison of their centrality rankings, which indicates that their structures are all but identical. A visual inspection of their respective rankings in Figure~\ref{fig:afl2_centrality_scatterplots} reveals no major differences, and the Kendall Tau and Spearman's coefficients indicate their rankings are, in fact identical (Figure~\ref{fig:afl2-centrality_ranking_comparisons_tau_rho}). 

\begin{table}[t]
    \centering\scriptsize
    \begin{tabular}{@{}rrr@{}}
        \toprule
        RETWEET & MENTION & REPLY \\
        \midrule
        0.916   &  0.808  & 0.865 \\
        \bottomrule
    \end{tabular}
    \caption{Adjusted Rand index scores for the clusters found in the networks built from the RAPID and Twarc datasets for the AFL2 collection.}
    \label{tab:afl2_ari_scores}
\end{table}

Interestingly, the high degree of similarity does not extend to the membership of detected clusters using the ARI measure (Table~\ref{tab:afl2_ari_scores}).
Presumably, the sensitivity of the measure indicates that these scores must be as close to the maximum as we could expect 
due to the degree of randomness inherent in Louvain clustering.
For a score of $1.0$ each pair of detected clusters would need to match perfectly, across the thousands of nodes in the networks, so any minor variation will reduce that score.

\subsubsection{Summary of findings}

This evidence suggests that the results provided by the Twitter API (if not other platforms' APIs) are consistent, regardless of the consumer. 
It is clearly important that a researcher understand how their collection tool works to guarantee their understanding of the results returned. In this regard, open source solutions are, as the name implies, more transparent than closed source solutions. The benefit gained as a result of more tailored filtering must be balanced against the initial effort required to understand how the APIs are employed by the tool used and what modifications tools make to the data they collect.



\subsection{Case Study 4: Election Day}\label{sec:cs4_election}

\begin{figure}[!ht]
    \centering
    \includegraphics[width=\columnwidth]{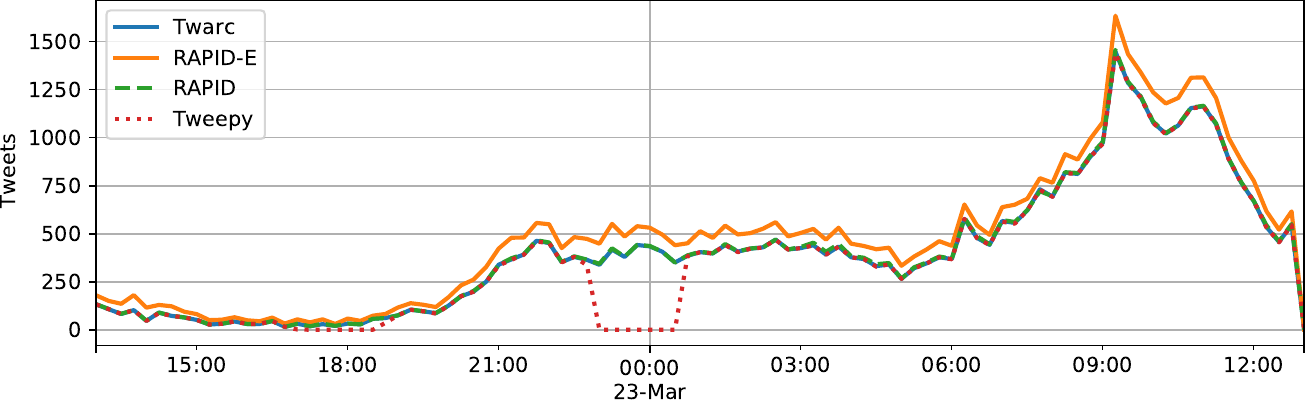}
    \caption{Twitter activity in the Election Day dataset over time (in 15 minute blocks). Dashed and dotted lines are used here to highlight how the timeseries overlap almost exactly.}
    \label{fig:nswelec_timeline}
\end{figure}

A final case study highlights the importance of continuous network connectivity, and awareness of when that condition is not met. 
To consider a more focused collection activity and to consider a second open source collection tool (thus similar to the baseline tool, Twarc), 
a collection was conducted over an election day (24 hour period) in early 2019, using RAPID, Twarc and Tweepy, 
each configured with the same filter terms: \texttt{\#nswvotes}, \texttt{\#nswelection}, \texttt{\#nswpol}, and \texttt{\#nswvotes2019}.  
RAPID and Twarc collected slightly below 40k tweets each while Tweepy collected around 36k tweets, but suffered from network outages on two occasions for approximately $110$ and $96$ minutes each time (see Figure \ref{fig:nswelec_timeline}). In the resulting datasets (highlights of which are shown in Table~\ref{tab:nswelec_collection_stats}), $285$ tweets were unique to RAPID, three to Twarc, and $19$ were shared by Twarc and Tweepy but not RAPID. The vast majority of the Tweepy dataset's $36,172$ tweets appeared in all three datasets, while Tweepy missed the $3,118$ further tweets that appeared in both Twarc and RAPID datasets. In fact, by examining the periods where Tweepy lost its connection, around 6~p.m. (UTC) and again approximately six hours later, Twarc retained $3,036$ tweets while RAPID retained $3,055$ tweets (RAPID-E collected $3,918$ during these periods), so it is possible that if Tweepy's connection had stayed up, the Tweepy dataset might have been very similar to Twarc and RAPID, especially as the remainder of the collection behaviour of the tools appears almost identical in the timeline.  


\begin{table}[t]
    \centering
    \resizebox{\columnwidth}{!}{%
        \begin{tabular}{@{}l|r|rr|rr|r|rr@{}}
        \toprule
        Dataset & \multicolumn{1}{c}{All} & \multicolumn{2}{c}{Unique} & \multicolumn{2}{c}{Retweets} & \multicolumn{1}{c}{All} & \multicolumn{2}{c}{Unique} \\ 
                & \multicolumn{1}{c}{Tweets} & \multicolumn{2}{c}{Tweets} & \multicolumn{2}{c}{} & \multicolumn{1}{c}{Accounts} & \multicolumn{2}{c}{Accounts} \\
        \midrule
        Twarc   & 39,293 & 3      & (0.0\%)  & 26,412   & (67.2\%) & 10,860   & 1        & (0.0\%)  \\
        RAPID   & 39,556 & 285    & (0.7\%)  & 26,612   & (67.3\%) & 10,893   & 36       & (0.3\%)  \\
        RAPID-E & 46,526 & 7,255  & (15.6\%) & 30,735   & (66.1\%) & 12,696   & 1,839    & (14.5\%) \\
        Tweepy  & 36,172 & 0      & (0.0\%)  & 24,276   & (67.1\%) & 10,242   & 0        & (0.0\%)  \\
        \bottomrule
        \end{tabular}%
    }
    \caption{Summary dataset statistics of the Election Day collection.}
    \label{tab:nswelec_collection_stats}
\end{table}

\subsubsection{Comparison of collection statistics}

The collection statistics are highly similar and are provided primarily for completeness. The effect of Tweepy's disconnection is highlighted by the differences in its statistics from Twarc as the baseline. Although more than $3,000$ tweets were missed, only a few hundred accounts, quotes, replies and tweets with URLs were missed. Several thousand retweets were missed as well as tweets with hashtags and mentions, but the effect on the features with the highest counts is limited. The most prolific account, most retweeted tweet, most replied to tweet, most mentioned accounts, hashtags and URLs are all the same.

\begin{table}[ht]
    \centering
    \resizebox{\columnwidth}{!}{%
    \begin{tabular}{@{}lrrr@{}}
        \toprule
        Property                           & Twarc                   & RAPID                   & Tweepy                  \\
        \midrule
        Tweets                             &                  39,297 &                  39,556 &                  36,172 \\
        \midrule
        Accounts                           &                  10,860 &                  10,893 &                  10,242 \\
        Retweets                           &                  26,412 &                  26,612 &                  24,276 \\
        Quotes                             &                   3,590 &                   3,610 &                   3,363 \\
        Replies                            &                   1,374 &                   1,381 &                   1,252 \\
        Tweets with hashtags               &                  21,582 &                  21,686 &                  19,977 \\
        Tweets with URLs                   &                   7,829 &                   7,860 &                   7,194 \\
        Most prolific account              & Account $a_8$           & Account $a_8$           & Account $a_8$           \\
        Tweets by most prolific account    &                     212 &                     211 &                     212 \\
        Most retweeted tweet               & Tweet $t_9$             & Tweet $t_9$             & Tweet $t_9$             \\
        Most retweeted tweet count         &                     368 &                     367 &                     278 \\
        Most replied to tweet              & Tweet $t_{10}$          & Tweet $t_{10}$          & Tweet $t_{10}$       \\
        Most replied to tweet count        &                      25 &                      26 &                      24 \\
        Tweets with mentions               &                  30,626 &                  30,848 &                  28,154 \\
        Most mentioned account             & Account $a_9$           & Account $a_9$           & Account $a_9$           \\
        Mentions of most mentioned account &                   2,442 &                   2,443 &                   2,187 \\
        Hashtag uses                       &                  51,288 &                  51,470 &                  47,106 \\
        Unique hashtags                    &                   2,450 &                   2,458 &                   2,306 \\
        Most used hashtag                  & \texttt{\#nswvotes}     & \texttt{\#nswvotes}     & \texttt{\#nswvotes}     \\
        Most used hashtag count            &                  11,739 &                  11,731 &                  10,901 \\
        Next most used hashtag             & \texttt{\#nswvotes2019} & \texttt{\#nswvotes2019} & \texttt{\#nswvotes2019} \\
        Uses of next most used hashtag     &                   7,606 &                   7,602 &                   6,968 \\
        URL uses                           &                   3,766 &                   3,761 &                   3,478 \\
        Unique URLs                        &                   1,374 &                   1,374 &                   1,258 \\
        Most used URL                      & URL 1\ts{*}             & URL 1\ts{*}             & URL 1\ts{*}             \\
        Uses of most used URL              &                     100 &                     100 &                     100 \\
        \bottomrule
    \end{tabular}
    } 
    \caption{Statistics of the Twarc, RAPID, and Tweepy datasets collected in parallel over a $24$ hour period. \ts{*}\texttt{https://www.fiverr.com/s2/ee030ef08d}}
    \label{tab:nswelec_all_collection_stats}
\end{table}

\subsubsection{Comparison of network statistics}

\begin{figure}[!ht]
    \centering
    \includegraphics[width=0.99\columnwidth]{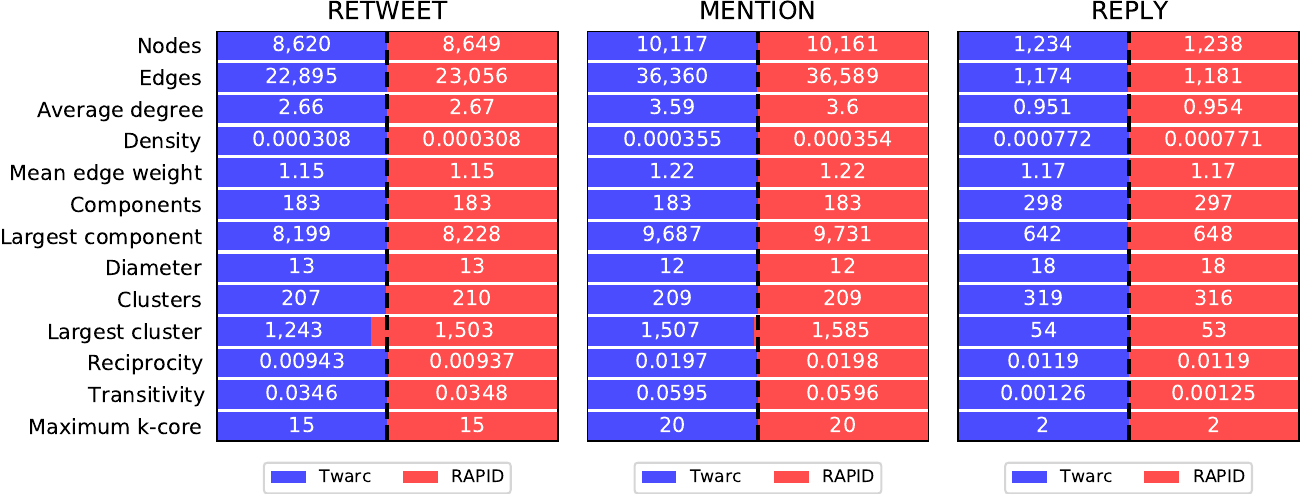}
    \caption{The proportional balance between Twarc and RAPID statistics of the retweet, mention and reply networks built from the Twarc and RAPID datasets.}
    \label{fig:nswelec-tvr_graph_stats_barh}
\end{figure}

\begin{figure}[!ht]
    \centering
    \includegraphics[width=0.99\columnwidth]{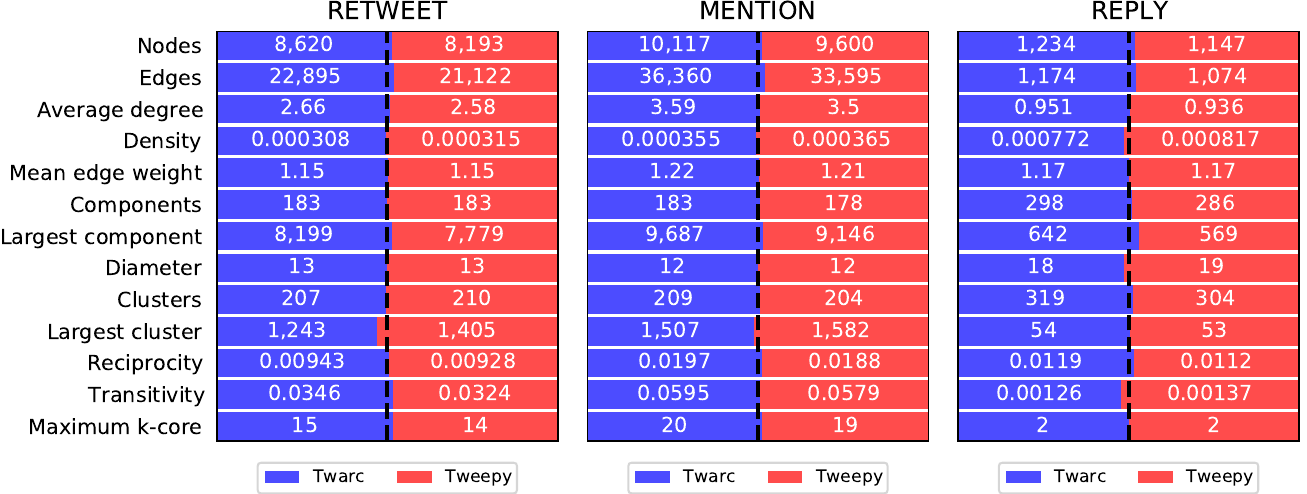}
    \caption{The proportional balance between Twarc and RAPID statistics of the retweet, mention and reply networks built from the Twarc and Tweepy datasets.}
    \label{fig:nswelec-tvt_graph_stats_barh}
\end{figure}

Continuing the similarities in the collection statistics, statistics drawn from retweet, mention and reply networks built from the Election Day datasets are also strikingly resilient, despite the Tweepy networks including several hundred fewer nodes (Table~\ref{tab:nswelec_graph_stats}). This is borne out by the proportional differences between Twarc and RAPID in 
Figure~\ref{fig:nswelec-tvr_graph_stats_barh}, where the only significant difference is the size of the largest detected cluster (again, likely due to the randomness inherent in Louvain~\citep{blondel2008louvain}), and then in the proportional differences in all the statistics across the Twarc and Tweepy networks in 
Figure~\ref{fig:nswelec-tvt_graph_stats_barh}.

\begin{table}[t]
    \centering
    \resizebox{\columnwidth}{!}{%
    \begin{tabular}{@{}lrrrrrrrrr@{}}
        \toprule
                           & \multicolumn{3}{c}{RETWEET}    & \multicolumn{3}{c}{MENTION}    & \multicolumn{3}{c}{REPLY}      \\ 
                           & Twarc    & RAPID    & Tweepy   & Twarc    & RAPID    & Tweepy   & Twarc    & RAPID    & Tweepy   \\
                           \cmidrule(lr){2-4}            \cmidrule(lr){5-7}            \cmidrule(lr){8-10}
        Number of nodes    & 8,620    & 8,649    & 8,193    & 10,117   & 10,161   & 9,600    & 1,234    & 1,238    & 1,147    \\ 
        Number of edges    & 22,895   & 23,056   & 21,122   & 36,360   & 36,589   & 33,595   & 1,174    & 1,181    & 1074     \\
        Average degree     & 2.656    & 2.666    & 2.578    & 3.594    & 3.601    & 3.500    & 0.951    & 0.954    & 0.936    \\
        Density            & 0.000    & 0.000    & 0.000    & 0.000    & 0.000    & 0.000    & 0.001    & 0.001    & 0.001    \\
        Mean edge weight   & 1.154    & 1.154    & 1.149    & 1.217    & 1.218    & 1.209    & 1.170    & 1.169    & 1.166    \\
        Component count    & 183      & 183      & 183      & 183      & 183      & 178      & 298      & 297      & 286      \\
        Largest component  & 8,199    & 8,228    & 7,779    & 9,687    & 9,731    & 9,146    & 642      & 648      & 569      \\
        - Diameter         & 13       & 13       & 13       & 12       & 12       & 12       & 18       & 18       & 19       \\
        Clusters           & 207      & 210      & 210      & 209      & 209      & 204      & 319      & 316      & 304      \\
        Largest cluster    & 1,243    & 1,503    & 1,405    & 1,507    & 1,585    & 1,582    & 54       & 53       & 53       \\
        Reciprocity        & 0.009    & 0.009    & 0.009    & 0.020    & 0.020    & 0.019    & 0.012    & 0.012    & 0.011    \\
        Transitivity       & 0.035    & 0.035    & 0.032    & 0.060    & 0.060    & 0.058    & 0.001    & 0.001    & 0.001    \\
        Maximum k-core     & 15       & 15       & 14       & 20       & 20       & 19       & 2        & 2        & 2        \\
        \bottomrule
    \end{tabular}
    } 
    \caption{Comparative statistics for networks generated from the Twarc, RAPID and Tweepy datasets for the Election Day collection.}
    \label{tab:nswelec_graph_stats}
\end{table}


\subsubsection{Comparison of centralities}

\begin{figure}[!ht]
    \centering
    \includegraphics[width=0.99\columnwidth]{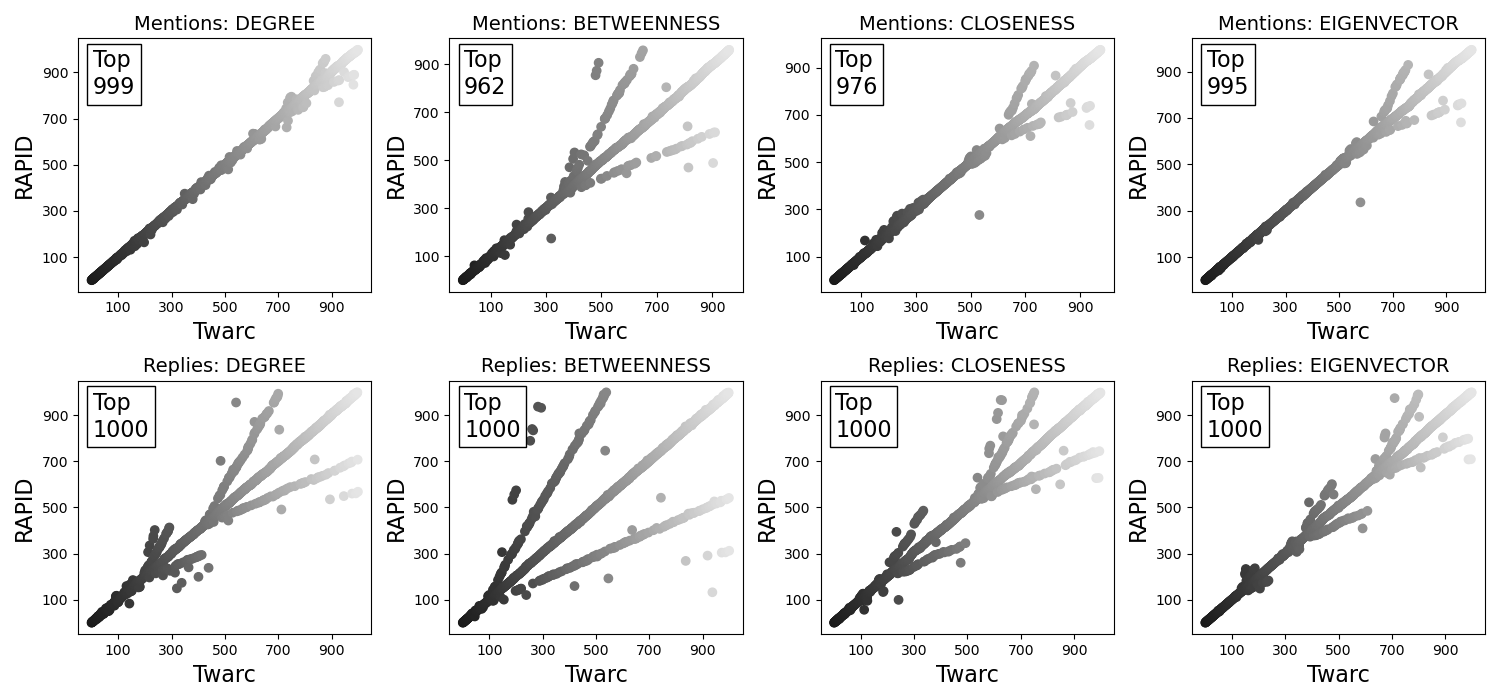}
    \caption{Centrality  ranking  comparison  scatter  plots  of  the  mention  and  reply  networks built from the Twarc and RAPID Election Day datasets. In each plot, each point represents a node’s ranking in the RAPID1 and RAPID2 lists of centralities (common nodes amongst the top $1,000$ of each list). The number of nodes appearing in both lists is inset. Point darkness indicates rank on the $x$ axis (darker = higher).}
    \label{fig:nswelec_tvr_centrality_scatterplots}
\end{figure}

\begin{figure}[!ht]
    \centering
    \includegraphics[width=0.99\columnwidth]{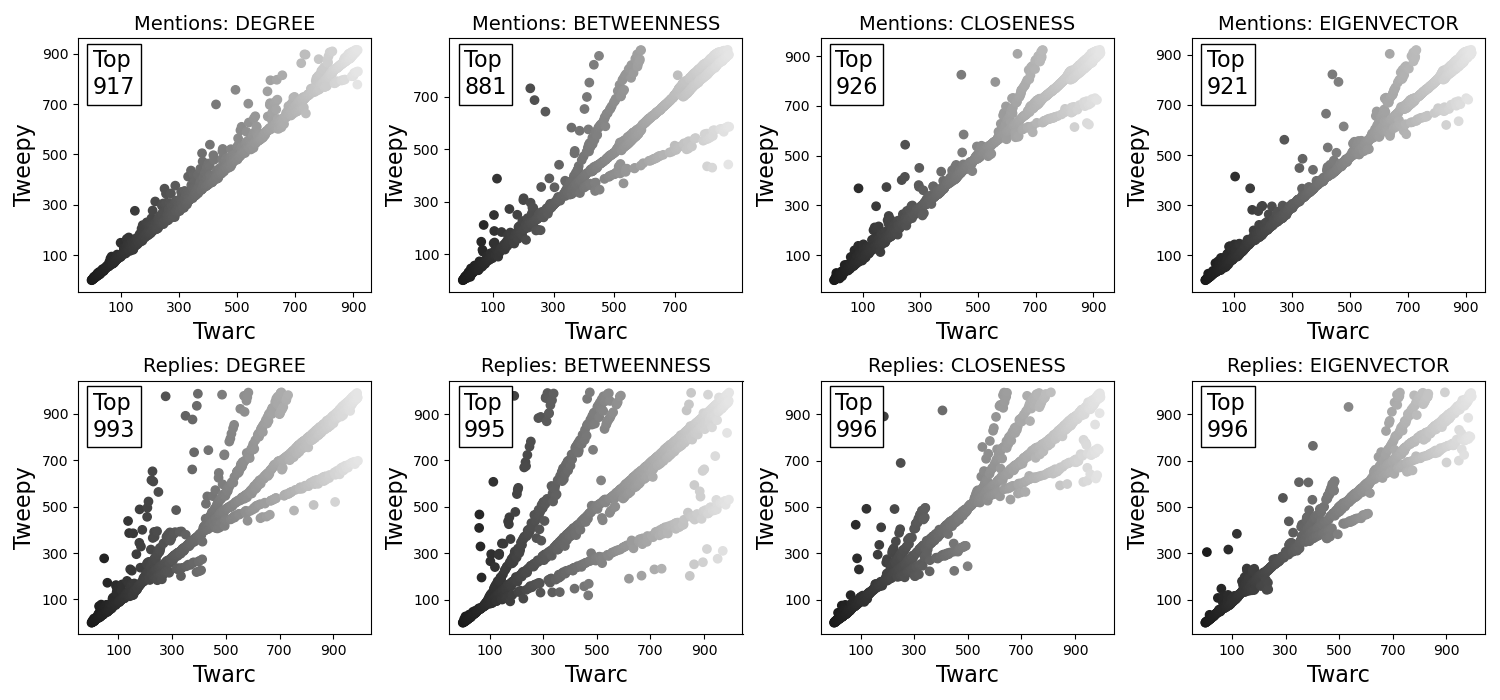}
    \caption{Centrality  ranking  comparison  scatter  plots  of  the  mention  and  reply  networks built from the Twarc and Tweepy Election Day datasets. In each plot, each point represents a node’s ranking in the RAPID1 and RAPID2 lists of centralities (common nodes amongst the top $1,000$ of each list). The number of nodes appearing in both lists is inset. Point darkness indicates rank on the $x$ axis (darker = higher).}
    \label{fig:nswelec_tvt_centrality_scatterplots}
\end{figure}

\begin{figure}[t!]
    \centering
    \includegraphics[width=0.99\columnwidth]{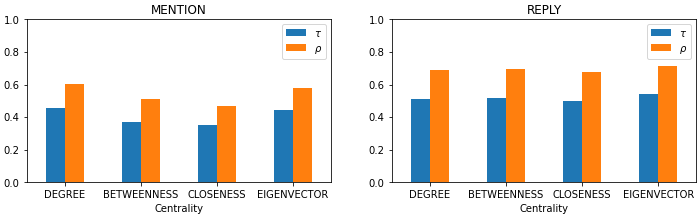}
    \caption{Centrality ranking comparisons from the Twarc and RAPID datasets of the Election Day collection using Kendall Tau scores and Spearman's Coefficients.}
    \label{fig:nswelec-tvr-centrality_ranking_comparisons_tau_rho}
\end{figure}

\begin{figure}[t!]
    \centering
    \includegraphics[width=0.99\columnwidth]{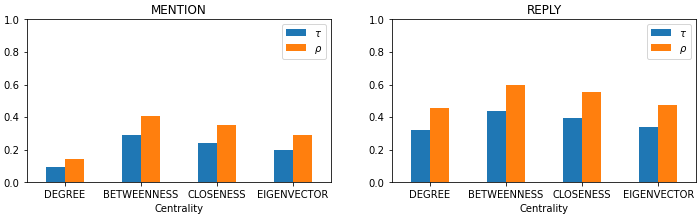}
    \caption{Centrality ranking comparisons from the Twarc and Tweepy datasets of the Election Day collection using Kendall Tau scores and Spearman's Coefficients.}
    \label{fig:nswelec-tvt-centrality_ranking_comparisons_tau_rho}
\end{figure}

Examining the centralities of the mention and reply networks built from the Election Day datasets, comparing RAPID and Tweepy against the Twarc baseline shows, as expected, only minor variations in the RAPID dataset which only occur among the lower ranked nodes (Figure~\ref{fig:nswelec_tvr_centrality_scatterplots}) and more widespread differences with the Tweepy networks (Figure~\ref{fig:nswelec_tvt_centrality_scatterplots}). Statistically, Twarc and RAPID's mention network centrality rankings, shown in Figure~\ref{fig:nswelec-tvr-centrality_ranking_comparisons_tau_rho}, had Kendall $\tau$  values around $0.35$ to $0.4$ and Spearman's coefficients around $0.45$ to $0.6$, while the reply networks' values were higher, with $\tau$ around $0.5$ and Spearman's coefficient around~$0.7$, possibly due to the smaller size of the reply networks. These values are all approaching or exceeding the $\tau$ value of $0.4$ to $0.6$ that was regarded as reasonably to highly similar, mentioned in Section~\ref{sec:analyses}. The ranking similarity statistics calculated by comparing the Twarc and Tweepy baselines are notably lower (Figure~\ref{fig:nswelec-tvt-centrality_ranking_comparisons_tau_rho}, though even the reply networks' betweenness and closeness comparisons are moderately similar with $\tau$ around $0.4$ and Spearman's coefficient around $0.5$ to $0.6$.

\subsubsection{Comparison of clusters}

Despite the similarities between the Twarc and RAPID networks, the cluster membership still varies significantly, with the highest similarity being found amongst the (smaller) reply networks, as can be seen in the ARI scores in Table~\ref{tab:nswelec_ari_scores}. The clusters found in the Twarc and Tweepy networks are less similar, almost in line with the differences in network sizes: the retweet networks had fewer nodes than the mention networks, and the ARI scores are less different, and the reply networks were the smallest and had the smallest difference between ARI scores. 

\begin{table}[ht]
    \centering\small
    \begin{tabular}{@{}lrrr@{}}
        \toprule
                     & RETWEET & MENTION & REPLY \\
        \midrule
        Twarc/RAPID	 &   0.547 &   0.656 & 0.737 \\
        Twarc/Tweepy &   0.453 &   0.534 & 0.703 \\
        \bottomrule
    \end{tabular}
    \caption{Adjusted Rand index scores for the clusters found in the corresponding retweet, mention and reply networks built from the Election Day datasets.}
    \label{tab:nswelec_ari_scores}
\end{table}

\subsubsection{Summary of findings}

This final case study provides us with further confidence that the differences observed early on in the Q\&A datasets are primarily caused by enhancements provided by the RAPID platform and the differences in the AFL1 datasets were due, in part, to the choice of ``afl'' as the lone filter term. The Election Day collection used several specific filter terms and ran long enough to collect several tens of thousands of tweets, enough time to avoid minor differences in start and stop times. Even the differences that did occur did not result in significant effects on several networks constructed from the data or on network analysis measures calculated over those networks.

\section{Discussion}





The case studies presented highlight how decisions regarding collection specification, such as the filter terms used or their number, and the collection duration, can result in datasets that trigger features in complex collection tools, quite apart from configuration of such tools to dynamically change the collection specification (e.g., use of RAPID's topic tracking feature). The primary variations explored here involved filter terms although collection duration also varied, depending on 
the collection event. The biggest variations in parallel datasets appeared when few filter terms were used and when they were short (i.e., having few characters), resulting in incidental noise from posts in unexpected languages (\texttt{\#qanda}) or with unexpected acronyms and from elements in post metadata (\texttt{\#afl}). When multiple terms were used, and when those terms were not valid words in a language (e.g., variations on \texttt{\#nswelec}), the parallel datasets were much more similar. Although it might be common sense to encourage careful design of collection specifications, these case studies highlight the value in (and danger in not) being more specific, by dictating the language of posts required as well as using multiple filter terms.

When variations in datasets occurred, the extra tweets resulted in the introduction of new nodes (accounts) in retweet, mention and reply networks, the majority of which were located within the largest connected network components (relatively few appeared as new, independent components). This consistently reduced the density of the retweet, mention and reply networks, but rarely affected the diameter of the largest component (Q\&A Part 2's retweet network is an exception here), implying that the new nodes appear in the 
core 
of the components, rather than on the periphery. Consequently, the extra nodes increased reciprocity, transitivity, and sometimes maximum k-core values in retweet and mention networks, but rarely changed reply networks. Reply interactions occurred least frequently in all datasets, and so reply networks were the least different in raw size (nodes and edges). 

The effects of collection variation were most prominent in centrality scores, particularly when the collection event involved direct interaction between participants (e.g., issue- or theme-based discussions such as during Q\&A and over weekends of football) and less straightforward information dissemination (e.g., during an election campaign). The ranking of nodes by centrality varied most in the mention and reply networks of Q\&A Part~1, even though more than half the top thousand ranked nodes in each pair of parallel networks were the same (an average of $560.5$ for mentions and $991$ for replies). The forking patterns appearing in scatter plots imply the presence of groups of nodes with adjacent centrality rankings, which then swapped when new nodes were added, possibly through impacting the internal topology of the largest components in some way. Spearman's $\rho$ and Kendall $\tau$ correlation coefficients were consistently higher for reply networks than mention networks, possibly due to their smaller size. No particular patterns in differences between centrality types were observed, which implies the differences between  pairs of parallel networks did not result in significantly different topologies. 

A final lesson drawn from the use of ARI scores is that even clustering of highly similar (e.g., almost identical) networks (in Case Study~$4$), result in ARI scores around $0.7$, meaning that ARI scores around $0.4$ can be seen as confirmation cluster membership is, in fact, quite similar.

The central 
purpose 
of this paper is to draw attention to unexpected variations in datasets collected from social media streams and the networks constructed from them. This is especially relevant when it is known that the stream is limited (either through platform rate-limiting or through platform algorithms, as occurs with, say, Twitter's 1\% sample stream). An obvious follow-up question is whether or not an objective measure of reliability is feasible. This relates closely to the question of how representative samples provided by platforms are of their entire data holdings, e.g., as studied by ~\cite{morstatter2013sample,gonzalez2014assessing,JosephLC2014comparison}, but that question relies on examining the choices made by the platform in deciding what to include in the sample they offer. Here, similar to Paik \& Lin~\cite{PaikLin2015}, our interest is in confirming that the data we request from a platform (with filter terms) matches what it has, or is at least representative of what it has (if rate-limits are encountered). Such a measure might rely on comparing the distributions of various features in our result dataset and the overall dataset (held by the platform), such as the accounts and the number of tweets they post, the number of hashtags, URLs, and mentioned used and replies, quotes and retweets made. Only the platform has sufficient information to provide this measure, and there may be significant value in them doing it for free or low cost streams they provide to researchers, analysts and other social media mass consumers. Providing a measure of representativeness (indicating reliability) along with results could: 1) encourage consumers to pay for the higher cost streams, while also 2) providing consumers with more certainty in any conclusions they draw from the results they analyse. A measure of \emph{reliability} rather than \emph{representativeness} could be more useful also, because there may be good reason for the results of filters that the platform provides to not be truly representative -- this would be the case when the full dataset includes significant amounts of spam or pornography\footnote{This raises the question of how one would deliberately study such topics, however.}. The reliability measure would indicate how representative the provided results are compared with the full 
results.

\section{Conclusion}






Under a variety of conditions, the collection tools employed in several use cases provided different views of specific online discussions. 
These differences manifested as variations in collection statistics, and network-level and node-level statistics for retweet, mention and reply social networks built from the collected data. 
Extra tweets were most often collected by Twarc, and these appear to have resulted in more connections in the largest components without affecting their diameters. This may affect results of diffusion analysis, as reachability correspondingly increases. 
Deeper study of reply content is required to inform discussion patterns. 




How reliable social media can be as a source for research without deep knowledge of the effects of collection tools on analyses is an open question. 
If a tool \emph{adds value} through analytics, what is the nature of the effect? 
This paper provides a methodology to explore those effects. 
A canonical measure of the reliability of a dataset would be valuable to the research and broader social media analysis community. This measure would explain how complete the results of a search or filter of live posts is, and if it is not complete, how representative the provided sample results are of the complete results. Only the platforms have this information to hand, however there would be benefits for them to do so, including as an enticement to consumers to pay for greater access to platform data holdings, as well as helping inform consumers of the degree to which they can depend on analyses of the data they receive. 

We recommend the following to those using OSN data: 
\begin{itemize}
    \item 
    Be aware of tool biases and their effects.
    \item Take care to specify filter and search conditions with keywords that capture relevant data and avoid irrelevant data, and make use of metadata filters to avoid unwanted content, e.g., constraining language code. Beware of short filter terms and ones that are meaningful in non-target languages.

    
    

    \item Check the integrity of data. We observed gaps and minor inconsistencies in the Election case study 
    due to connection failures as well as the appearance of duplicate tweets, identical in data and metadata.
\end{itemize}

There are a number of avenues by which to expand this research:
\begin{itemize}
    \item Consolidate and expand the methodology so that it may be applied to other OSNs and collection tools, especially proprietary ones. A method to shed light on the biases of tools would be of great value to the community.
    
    \item Introduce content analysis to the comparison. If social networks built from parallel datasets vary, what is the effect on analyses of the discussion? Are text analysis methods robust enough to overcome differences or is it possible to draw entirely different conclusions, depending on the method used?
    
    \item Examine differences in information flow based on clusters rather than just individual nodes to help inform questions about broad information flow within the networks. If the differences are moderate, then we may draw confidence that overall flows of influence in a network may remain relatively steady, even with variations in the collection. 
    
    \item Although OSNs are similar when the interaction primitives they offer are considered, the way in which their feature sets are presented in user interfaces create a platform-specific interaction culture, and that affects the observable behaviour of its users, which may in turn affect analyses. An exploration of how platform culture manifests itself and differs would help inform the search for higher level social activities such as coordination~\citep{GrimmeAA2018perspectives,PachecoFM2020whitehelmets,weber2020coord,Graham2020}.
    
    \item Key for future social media research is to develop processes for repeatable analysis, including access to common datasets, both of which underpin the practice of benchmarking. Currently, OSN terms and conditions\footnote{\url{https://developer.twitter.com/en/developer-terms/agreement-and-policy#id8}} often hamper researchers exchanging datasets, so new techniques cannot easily be evaluated on the same data, raising questions of fair comparison. For example, Twitter requires that only tweet IDs are shared, forcing the next researcher to `re-hydrate' the tweets by downloading them again from Twitter's servers. By the time a new researcher does this, the data may have changed: metadata, such as retweet and like counts are constantly incremented; tweets and entire accounts may have been deleted or removed through suspension or account closure; or accounts may become private and inaccessible.
\end{itemize}

Finally: Does it matter if a streamed collection is not necessarily either complete or representative? As long as a researcher makes clear how they conducted a collection and using what tools and configuration, does it not still result in an analysis of behaviour that occurred online?
The answer is that it very much depends on the conclusions being drawn. Yes, the collection represents real activity that occurred, but it may not be complete, and conclusions drawn from it may be unintentionally misinformed and lacking in nuance. 
We have seen that variations in collections have an impact on network size and structure. This may result in different community compositions and affect centrality analyses, consequently misleading influential account identification and expected diffusion patterns. 
A firm understanding of the data and how it was obtained is therefore vital.

\section*{Declarations}

\subsection*{Funding} Mehwish Nasim acknowledges that this work has been partially supported by the Cyber Security Research Centre Limited whose activities are partially funded by the Australian Government's Cooperative Research Centres Programme.

\subsection*{Conflicts of interest/Competing interests} Not applicable.

\subsection*{Availability of data and material} The data (the identifiers of tweets only, as per Twitter's terms and conditions) used in this study are available at \url{https://github.com/weberdc/socmed_sna}.

\subsection*{Code availability} The data manipulation and analysis software written for this study is available at \url{https://github.com/weberdc/socmed_sna}.

\subsection*{Authors' contributions} All authors contributed to the study conception and design. Data collection was performed by Derek Weber, Lewis Mitchell, and Mehwish Nasim, and analysis was performed by Derek Weber, while advised by Mehwish Nasim, Lewis Mitchell and Lucia Falzon. The first draft of the manuscript was written primarily by Derek Weber with contributions from the other authors, and all authors commented on previous versions of the manuscript. All authors read, edited and approved the final manuscript.

\subsection*{Ethics approval} All data were collected, stored, processed and analysed according to two ethics protocols \#170316 and H-2018-045, approved by the University of Adelaide's human research and ethics committee.

\subsection*{Consent to participate} Not applicable.

\subsection*{Consent for publication} All authors consent to this work being published.

\bibliographystyle{spbasic}      
\bibliography{jsocnet}   


\end{document}